\documentclass[aps,pra,twocolumn,groupedaddress,amsmath,amssymb]{revtex4-1}

\usepackage{color}

\usepackage{bm}
\usepackage{nicefrac}
\usepackage[colorlinks,pdfusetitle,urlcolor=blue,citecolor=blue,linkcolor=blue]{hyperref}

\usepackage{graphicx}
\usepackage{braket}
\usepackage{math tools}
\newcommand{\pd}[2]{\frac{\partial#1}{\partial #2}}

\newcommand{\omegatrap}{\omega_{H}}

\newcommand{\akl}{\hat{a}_{\mathbf{k}\lambda}^{\vphantom{\dagger}}}
\newcommand{\akld}{\hat{a}_{\mathbf{k}\lambda}^{{\dagger}}}
\newcommand{\ax}{\hat{b}_x^{\vphantom{\dagger}}}
\newcommand{\axd}{\hat{b}_x^{{\dagger}}}
\newcommand{\ay}{\hat{b}_y^{\vphantom{\dagger}}}
\newcommand{\ayd}{\hat{b}_y^{{\dagger}}}
\newcommand{\aL}{\hat{b}_L^{\vphantom{\dagger}}}
\newcommand{\aLd}{\hat{b}_L^{{\dagger}}}
\newcommand{\aR}{\hat{b}_R^{\vphantom{\dagger}}}
\newcommand{\aRd}{\hat{b}_R^{{\dagger}}}
\newcommand{\ai}{\hat{b}_i^{\vphantom{\dagger}}}
\newcommand{\aid}{\hat{b}_i^{{\dagger}}}

\newcommand{\re}{\operatorname{Re}}

\newcommand{\B}[1]{\mathbf{#1}}
\newcommand{\SumOverModes}{\int d^3\B{k} \sum_{\lambda}}

\begin{document}


\title{Zeeman shift of an electron trapped near a surface}

\author{Robert Bennett}
\altaffiliation[Present address: ]{School of Physics and Astronomy, University of Leeds, Leeds LS2 9JT}
\author{Claudia Eberlein}
\affiliation{Department of Physics \& Astronomy, University of Sussex, Falmer, Brighton BN1 9QH, UK}


\date{\today}

\begin{abstract}
Boundary dependent corrections to the spin energy eigenvalues of an electron in a weak magnetic field and confined by a harmonic trapping potential are investigated. The electromagnetic field is quantized through a normal mode expansion obeying the Maxwell boundary conditions at the material surface. We couple the electron to this photon field and a classical magnetic field in the Dirac equation, to which we apply the unitary Foldy-Wouthuysen transformation in order to generate a non-relativistic approximation of the Hamiltonian to the desired order. We obtain the Schr\"{o}dinger eigenstates of an electron subject to double confinement by a harmonic potential and a classical magnetic field, and then use these within second-order perturbation theory to calculate the spin energy shift that is attributable to the surface-modified quantized field. We find that a pole at the eigenfrequency of a set of generalized Landau transitions gives dominant oscillatory contributions to the energy shift in the limit of tight harmonic confinement in a weak magnetic field, which also make the energy shift preferable to the magnetic moment for a physically meaningful interpretation.  \end{abstract}


\maketitle
\section{Introduction}
Progress in experimental physics over the last decade has meant that microscopic objects such as atoms or ions are now routinely trapped and manipulated in the laboratory. This has applications to nanotechnology as well as to ultra-precise tests of fundamental physics. The latter are now in some cases so precise that entirely new theoretical considerations have to be made in order to enumerate all possible systematic effects. One source of such effects can be the experimental apparatus itself, intentionally or unintentionally, because the quantized electromagnetic vacuum field is modified by the presence of macroscopic objects, while also being coupled to microscopic systems in the vicinity. This causes the properties of a microscopic system near a surface to differ from those found in free space, with the most famous example being the Casimir-Polder shift of the energy levels of a neutral atom near a surface \cite{CP}, and the surface-dependence of the anomalous magnetic moment of the electron \cite{Fischbach, Svozil, Bordag, BoulwareBrown, KreuzerSvozil, Kreuzer, BartonFawcett, BennettEberleinMagMomNJP, MagMomPRA} being another example.

In our previous work  \cite{BennettEberleinMagMomNJP,MagMomPRA, MassShiftsPaper} we calculated the mass and magnetic moment shifts for a free particle near a variety of surfaces via an explicit mode expansion of the quantized electromagnetic field. We showed that such surface-dependent radiative corrections can be calculated using a set of techniques from quantum optics, sidestepping many of the difficulties with a full field-theoretic approach, which is unnecessary as the calculation of leading-order corrections does not require the quantization of the matter field. 
Using perturbation theory in the Dirac equation, we derived general formulae which deliver the mass and magnetic moment shifts of an electron near an interface that is described by its reflection coefficients for transverse electric (TE) and transverse magnetic (TM) modes.
  
In this paper we extend our calculations of radiative corrections to an electron that is subject to a harmonic confining potential near a surface. On the one hand, this represents a more realistic model of the experimental conditions under which precision measurements of the properties of an electron are carried out, for example in a Penning trap \cite{GabrielsegfactorPRL}, and on the other hand, this can also be used as a model for describing an electron bound in an atom. We will derive the shift in the energy gap between the electron's two spin states in a weak homogeneous magnetic field; more precisely the part of that shift that is attributable to the presence of the surface. This is the same quantity from which we extracted the magnetic moment in our previous work \cite{BennettEberleinMagMomNJP,MagMomPRA}. We shall see, however, that in the presence of a trapping potential, the magnetic moment is not always a physically meaningful and measurable quantity and it is often more useful to discuss energy shifts instead.

Our starting point is the Dirac equation coupled to an electromagnetic field $A_\mu$
\begin{equation}
[-i\gamma^\mu (\partial_\mu + ie A_\mu) + m] \psi = 0.
\end{equation}
with the $\gamma$ matrices as defined in their standard Dirac representation. As we are interested in a non-relativistic expansion, we will work with the Dirac equation in its non-covariant form
\begin{equation}
i\frac{\partial }{\partial t} \psi = [\bm{\alpha} \cdot (\B{p}-e\B{A}) +e \Phi + \beta m] \psi \equiv H_D \psi
\label{Dirac}
\end{equation}
obtained by replacing $\gamma^0 = \beta$, $\gamma^i = \beta \alpha^i$ and $A_\mu=( \Phi, -\B{A})$. We subject the electron to a weak classical field $\mathbf{B}_0$ acting along the $\hat{z}$ axis; $\mathbf{B}_0 = B_0 \hat{z}$. A suitable classical vector potential is given by $\mathbf{A_0}=-\frac{1}{2} (\mathbf{r} \times \mathbf{B}_0 )$, to which we add the quantized Maxwell field $\mathbf{A}_q$, so that the total vector field that couples to the electron is $\mathbf{A} = \mathbf{A}_0 + \mathbf{A}_q$. We would like to take a non-relativistic approximation of Eq.~(\ref{Dirac}) right from the start, so that we can work with the Schr\"{o}dinger eigenstates of the electron. In previous calculations of the magnetic moment we have used the Dirac Hamiltonian $H_D$ directly within second-order perturbation theory using the Dirac eigenstates for an electron in a constant magnetic field \cite{JohnsonLippmann} but without any other confining potential. However, the Dirac eigenstates for an electron which is confined by a harmonic potential \emph{as well as} a constant magnetic field are not easily derivable from the solution of the Schr\"{o}dinger equation, because the square of the Dirac Hamiltonian with a potential $V({\bf r})$ can no longer be expressed just in terms of the corresponding Schr\"odinger Hamiltonian (cf.~App.~A of \cite{MagMomPRA}). Consequently, we begin this calculation  by taking the Foldy-Wouthuysen transformation \cite{FWOriginal} of the Dirac Hamiltonian, which will furnish us with the relevant Schr\"{o}dinger Hamiltonian to any desired order in the non-relativistic approximation. This procedure requires some care, since several successive applications of the transformation must be applied. The result to order $1/m^3$ is, in agreement with \cite{BartonFawcett},
\begin{equation} H_S \equiv H_0 + H_1 + H_2 \label{NonRelHamiltonian}\end{equation}
with 
\begin{subequations}
\label{FWHamiltonian} 
\begin{align}
H_0=& H_{\text{rad}} + \frac{\bm{\pi}^2}{2m}  - \frac{e}{2m} \bm{\sigma} \cdot \B{B}_0 + V_{\text{image}}  \\
H_1=&\frac{e^2}{2m} \B{A}_q^2 +\frac{e^3}{4m^3}\B{A}^2_q \bm{\sigma} \cdot \B{B}_0 \label{FirstOrderHam}  \\
 H_2=&- \frac{e}{m} \B{A}_q\cdot \bm{\pi}-\frac{e}{2m} \bm{\sigma}\cdot \B{B}_q \notag \\
 &\qquad \qquad + \frac{e}{8m^2}\bm{\sigma} \cdot ( {\bm{\pi} \times \B{E}_q - \B{E}_q \times \bm{\pi}}) \label{SecondOrderHam} \end{align} \end{subequations}
where $\B{E}_q = -\pd{\B{A}_q}{t}$ and $\B{B}_q = \nabla \times \B{A}_q$ are the electric and magnetic fields associated with the quantized vector potential, $\boldsymbol{\pi}=\B{p}-e\B{A}_0$ is the canonical momentum, and $ V_{\text{image}}$ is the electrostatic image potential of the electron. $H_0$ is the unperturbed Hamiltonian, $H_1$ and $H_2$ are the parts contributing in first-order and second-order perturbation theory, respectively. 

\begin{figure}[h!]
\includegraphics[width = 8cm]{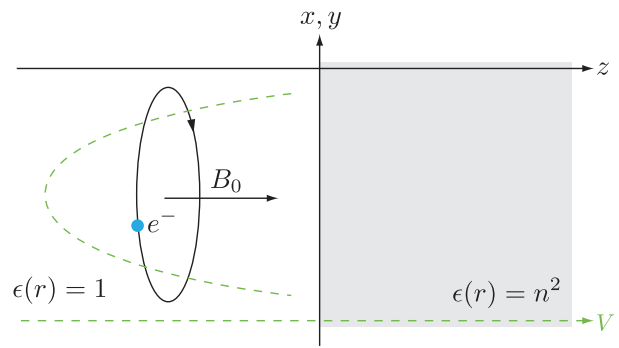}
\caption{\label{HalfSpace} Physical setup, with the horizontal axis representing the $z$ co-ordinate (solid lines) and the potential (dashed lines)}
\end{figure}

Perturbation theory applied to the Hamiltonian \eqref{NonRelHamiltonian} can be used to derive the spin-flip energy for a free electron close to a surface, from which a magnetic moment can be extracted with results in agreement with \cite{BennettEberleinMagMomNJP}. However, in this work we are interested in a harmonic confinement in the directions parallel to the surface, 
\begin{equation}
V_H=\frac{m \omegatrap^2}{2}(x^2+y^2)
\label{VH}
\end{equation}
as shown in Fig.~(\ref{HalfSpace}). Note that we do not consider confinement in the $z$ direction because this would not make any difference to the spin-flip energy in a magnetic field that is directed along $z$; the results for the magnetic moment shift are the same as those given in Refs.~\cite{BennettEberleinMagMomNJP,MagMomPRA} for an electron close of a surface in just a magnetic field. 

By contrast, for a potential in $x$ and $y$ as in Eq.~(\ref{VH}), there is interplay between the confinement due to $V_H$ and that due to the magnetic field ${\bf B}_0$ along $z$, which will be seen to lead to magnetic moment corrections different from those calculated in Refs.~\cite{BennettEberleinMagMomNJP,MagMomPRA} without any $V_H$.
The unperturbed Hamiltonian becomes
\begin{equation}
H_0^H = H_{\text{rad}} + \frac{\bm{\pi}^2}{2m}  - \frac{e}{2m} \bm{\sigma} \cdot \B{B}_0  +V_H \label{H0}
\end{equation}
and the energy shift up to second order in the perturbative expansion is
\begin{align}
\Delta E =\bra{\Psi_e, 0} H_1 &\ket{\Psi_e, 0}\notag \\
&+\sum_{\Psi_e'}  \frac{|\bra{\Psi_e', 1_{\B{k}\lambda}}H_2\ket{\Psi_e, 0}|^2}{E-E'}\label{PertExpansion}
\end{align}
where $ 1_{\B{k}\lambda}$ indicates a photon with wave vector $\B{k}$ and polarization $\lambda$, and $\Psi_e$ represents the state of the electron which is coupled to the classical field $B_0$ \emph{and} the confining potential  $V_H$.
The Schr\"odinger eigenstates $\Psi_e$ are determined in Appendix \ref{HOscApp}, where we find that Hamiltonian for an electron subject to this double confinement can be written as
\begin{equation}
H_e=\left(\Omega-\frac{eB}{2m}\right)\aRd \aR+\left(\Omega+\frac{eB}{2m}\right)\aLd \aL + \Omega
\end{equation}
where 
\begin{equation}
\Omega^2 =\omegatrap^2+\left(\frac{eB_0}{2m}\right)^2 \label{OmegaDef}
\end{equation}
and the operators $\aR$ ($\aRd$) and $\aL$ ($\aLd$) are the lowering (raising) operators for right and left-circular quanta in a set of generalized Landau levels labelled by quantum numbers $\nu_R$ and $\nu_L$, which we write as the composite state $\ket{\nu_L}\otimes\ket{\nu_R} = \ket{\nu_L,\nu_R}$ Combining this with the spin eigenstate $s$, we have 
\begin{equation}
\ket{\Psi_e} = \ket{\nu_L,\nu_R}\otimes \ket{s} = \ket{\nu_L,\nu_R; s}\; . 
\end{equation}

The electromagnetic field is written in terms of mode functions $\mathbf{f}_{\mathbf{k}\lambda}$ \cite{fn1}
%
\begin{equation}
\mathbf{A} = \sum_{\lambda} \int d^3 \B{k} \frac{1}{\sqrt{2 \omega}}(\mathbf{f}^{\vphantom{\dagger}}_{\mathbf{k}\lambda} \akl+\mathbf{f}_{\mathbf{k}\lambda}^* \akld) \; ,  \label{AField}
\end{equation}
where $\akl$ and $\akld$ are creation and annihilation operators for photons of wavenumber $k=|\B{k}|$ and polarization $\lambda$.  The mode functions for the quantized field near a non-dispersive dielectric are given in appendix \ref{ModesAppendix}. They are normalized (cf.~Eq.~(\ref{normf})) so that the Hamiltonian for the radiation field is mapped into the canonical form 
\begin{equation}
H_\text{rad} = \sum_\lambda \int d^3 \B{k} \, \omega_k \left(\akld \akl+\frac{1}{2}\right) \; . \label{Hrad}
\end{equation} 
We now use second-order perturbation theory to derive the shift in the Zeeman energy gap between the two spin states of the electron that is attributable to the presence of the surface which reflects and refracts the quantized electromagnetic field. In order to find all contributing terms, it is important to include next-to-leading order terms in the multipole expansion of the quantized field $\B{A}_q$:
\begin{equation}
\mathbf{A}_q(\mathbf{r}) = \mathbf{A}_q(\mathbf{r}_0) + [(\mathbf{r}-\mathbf{r}_0)\cdot\nabla]\mathbf{A}_q(\mathbf{r_0})+...\label{TaylorExp}
\end{equation}
As already explained in \cite{MagMomPRA}, the fact that this calculation requires the inclusion of terms beyond the usual dipole approximation is due to the curvature of classical trajectories which plays a role in some of the terms. 
The non-zero matrix elements of the displacement operator $\mathbf{r}-\mathbf{r}_0$ in the unperturbed eigenstates $\ket{\nu_L,\nu_R}$ are given by Eqs.~\eqref{PositionMatrixElements}. We note that the operator changes the Landau level $\nu_R$ or $\nu_L$, meaning that it can result in transitions to or from an intermediate state $\Psi_e'$ in second-order perturbation theory, thus contributing to the energy shift. 

\section{Perturbation Theory}
\subsection{First Order}
The first-order term in the perturbation expansion \eqref{PertExpansion} is
\begin{equation}
\Delta E_1 =\bra{\Psi_e, 0} H_1 \ket{\Psi_e, 0} \label{FirstOrderPert}
\end{equation}
Inserting  $H_1$ as shown in Eq.~\eqref{FirstOrderHam} into Eq.~\eqref{FirstOrderPert}, we have
\begin{align}
\Delta E_1=\frac{e^2}{2m}&\bra{\Psi_e, 0}  \left[ \B{A}_q^2 +\frac{e}{2m^2}\B{A}^2_q \bm{\sigma} \cdot \B{B}_0 \right] \ket{\Psi_e, 0} 
\end{align}
The first term is independent of $\bm{\sigma}$ and hence cannot contribute to a shift in the Zeeman energy gap between the two spin states; therefore we discard it. Throughout this paper we shall discard any such terms that shift all spin levels equally and thus do not change the energy gap between the two spin states. We are interested only in the part of the energy shift $\Delta E$ that affects the two spin states differentially, as this is the part that is spectroscopically accessible, and we shall denote that part by $\Delta \mathcal{E}$.

Using the expression given in Eq.~\eqref{AField} for the quantized electromagnetic field in terms of mode functions, we find for the remaining term \footnote{Energy shifts depend on the state being spin-up or spin-down. Here and throughout this paper we abbreviate this dependence by writing energy shifts as proportional to the Pauli spin matrix $\sigma_z$ which should be understood as a shorthand for the expectation value $\braket{ s| \sigma_z | s } $ .}
\begin{align}
&\Delta \mathcal{E}_1=\frac{e^3}{4m^3 } {B}_0 \bra{0,\nu_L,\nu_R,s} \sigma_z  \B{A}_q^2\ket{0,\nu_L,\nu_R,s} \notag \\
&=\frac{e^3}{8m^3 }\sigma_z {B}_0\int d^3 \B{k} \sum_{\lambda}\frac{1}{\omega} \left( |f_x|^2+|f_y|^2+|f_z|^2 \right) \label{FirstOrderContribution} \;.
\end{align}

\subsection{Second Order}
The second-order term in the perturbation expansion \eqref{PertExpansion} for the energy shift is
\begin{align}
\Delta E_2= \sum_{\Psi_e'}  \frac{|\bra{\Psi_e', 1_{\B{k}\lambda}}H_2\ket{\Psi_e, 0}|^2}{E-E'} \label{SecondOrderPertOnly}
\end{align}
with $H_2$ given by Eq.~\eqref{SecondOrderHam}. 
Noting that
\begin{eqnarray}
 \bm{\pi} \times \B{E}_q - \B{E}_q \times \bm{\pi}&=&-i(\bm{\nabla} \times \B{E}_q) - 2\B{E}_q \times \bm{\pi} \nonumber \\
 &=&-i\frac{\partial\B{B}_q}{\partial t} - 2\B{E}_q \times \bm{\pi}\;, \nonumber
\end{eqnarray}
we choose to split $H_2$ into
\begin{equation}
H_2 = H_2^E+ H_2^B
\end{equation}
with
\begin{equation}
H_2^E=- \frac{e}{m}\left[ \B{A}_q +\frac{1}{4m}  (\bm{\sigma} \times \B{E}_q) \right]\cdot \bm{\pi}\label{H2E} 
\end{equation}
and
\begin{equation}
H_2^B= -\frac{e}{2m} \bm{\sigma}\cdot \B{B}_q +\frac{ie}{8m^2}\bm{\sigma} \cdot \pd{\B{B}_q}{t}\;.  \label{H2B}
\end{equation}
In the dipole approximation, where the field operators $\B{A}_q$, $\B{E}_q$ and $\B{B}_q$ depend solely on ${\bf r}_0$ but not on ${\bf r}$ and hence cannot lead to a change in the Landau levels, $H_2^E$ in Eq.~\eqref{H2E} effectively contributes to the second-order shift in Eq.~\eqref{SecondOrderPertOnly} only for $s=s'$, because contributions of intermediate states with $s\neq s'$ would require terms with two $\bm{\sigma}$ matrices and thus be of next-to-leading order $\propto 1/m^4$ in the non-relativistic expansion. Therefore, in the dipole approximation and to leading order $\sim 1/m^3$ in the non-relativistic approximation, $H_2^E$ changes only the Landau level $\{ \nu_R, \nu_L \}$, and its contribution to the energy shift is
\begin{align}
\Delta  E_2^E &=\frac{e^2}{m^2} \sum_{\nu_R',\nu_L'} \notag \\
&\times
 \frac{| \bra{\nu_R',\nu_L'; 1_{\B{k},\lambda}}\left(\B{A}_q + \frac{\bm{\sigma} \times \B{E}_q}{4m}\right)\cdot \bm{\pi}\ket{\nu_R,\nu_L; 0}|^2}{-\omega + E_{\nu_L\vphantom{'},\nu_R\vphantom{'}}-E_{\nu_L',\nu_R'}}
\label{DeltaE2}
\end{align}
where the Schr\"odinger eigenvalue $E_{\nu_L\vphantom{'},\nu_R\vphantom{'}}$ is given by Eq.~\eqref{EnLnR}.
Defining 
\begin{equation}
\Delta_i \equiv \Omega -h_i\frac{eB_0}{2m}\label{DeltaDef}
\end{equation}
where $h_i$ denotes the handedness of the Landau states via
\begin{equation}
h_R=+1 \qquad h_L = -1\label{hDef}
\end{equation} 
and a generalized summation symbol
\begin{equation}
\widetilde{\sum} \equiv  \int d^3\B{k} \sum_{\lambda = \text{TE,TM}}\;  \sum_{i=\text{L},\text{R}}=\SumOverModes \sum_{i=\text{L},\text{R}}
\end{equation}
we can evaluate the four contributions to the sum over Landau levels ($\nu_R' = \nu_R \pm 1$, $\nu_L' = \nu_L \pm 1$) and extract the part of the energy shift \eqref{DeltaE2} that shifts the spin-up and spin-down states differentially, and obtain
\begin{align}
\Delta \mathcal{E}_2^E=-\frac{e^2}{16m^2}\sigma_z \widetilde{\sum} &\frac{\Delta_i^2 h_i}{\Omega}(|f_x|^2+|f_y|^2)  \notag \\
& \times \left( \frac{\Delta_i(2\nu_i+1)-\omega}{\omega^2-\Delta_i^2}\right) \label{DeltaEDip}
\end{align}

Moving on to $H_2^B$ in Eq.~\eqref{H2B}, which is the part that in the dipole approximation changes only the spin $s$ but not the Landau levels, 
we now calculate its contribution to the energy shift,
\begin{align}
\Delta  E_2^B  = \sum_{s'} \frac{\bra{s';  1_{\B{k},\lambda}}H_2^B \ket{s; 0}|^2}{-\omega + E_{s}-E_{s'}} 
\end{align}
where $E_s$ is the unperturbed Zeeman energy of the spin state,
\begin{align}
E_{s} &= -\frac{e B_0}{m}s\ , & s=\pm \frac{1}{2}
\end{align}
Inserting the explicit form of $H_2^B$ from Eq.~\eqref{H2B}, we find
\begin{align}
\Delta  E_2^B =\frac{e^2}{4m^2} & \int d^3 \B{k} \sum_{\lambda } \left( 1+\frac{\omega}{4m} \right)^2 \frac{1}{2\omega}  \notag \\
&\times \sum_{s'} \frac{| \bra{s'}\bm{\sigma} \cdot ( \nabla\times  \B{f})\ket{s}|^2}{-\omega -E_{s'} +E_{s}} \; . \label{DeltaESpin}
\end{align}
The second line of Eq.~\eqref{DeltaESpin} can be written as
\begin{align}
&\frac{1}{\omega}\frac{1}{\left( \frac{eB_0}{m}\right)^2-\omega^2 } \sum_{s'} \braket{s | \bm{\sigma} \cdot (\nabla \times \B{f})^* |s'}\braket{s' | \bm{\sigma} \cdot (\nabla \times \B{f}) |s} \notag  \\
&\times \left[ \omega + \frac{eB_0}{m} (s+s') \right] \left(\omega- \frac{2eB_0}{m} s\right) \label{DeltaESpinMatrixElement}\; . 
\end{align}
This can be readily simplified by re-writing the spin eigenvalue as an operator acting on the respective state, $s \ket{s} = \sigma_z/2\ket{s}$ and using the completeness of the spin states, $\sum_{s'} \ket{s'} \bra{s'} = \mathbb{I}$. Then, successively multiplying out all $\bm\sigma$ matrices and extracting the terms proportional to $\sigma_z$, we find
\begin{align*}
\Delta  \mathcal{E}_2^B =& -\sigma_z \frac{e^3B_0}{8m^3} \int d^3 \B{k} \sum_{\lambda}\left(1+\frac{\omega}{4m} \right)^2 \frac{1}{\omega} \\
&\times \frac{1}{\left( \frac{eB_0}{m}\right)^2-\omega^2 } \left[|(\nabla\times \B{f}^*)_x|^2+|(\nabla\times \B{f}^*)_y|^2\right]\;. \end{align*}
To the $1/m^3$ accuracy of the Foldy-Wouthuysen transformation carried out to find the Hamiltonian, this is:
\begin{equation}
\Delta  \mathcal{E}_2^B =\frac{e^3B_0\sigma_z}{8m^3} \int d^3 \B{k} \sum_{\lambda} \frac{1}{\omega^3}
\left[|(\nabla\times \B{f}^*)_x|^2+|(\nabla\times \B{f}^*)_y|^2\right] \label{DeltaESpinTermsOfF} \end{equation}

We have now found all the terms in the energy shift that, in the dipole approximation (i.e.~the leading term in Eq.~\eqref{TaylorExp}), are proportional to $\sigma_z$ and thus shift the two spin states differentially. To check whether any terms contribute beyond the dipole approximation, we do a multipole expansion of each term of $H_2$ as given by Eq.~\eqref{SecondOrderHam} via Eq.~$\eqref{TaylorExp}$ and indeed find two additional contributions. The first of these stems from application of the multipole operator to the term in $\bm{\sigma} \cdot \B{B}_q$,
 \begin{widetext}
\begin{equation}
\Delta E^{\text{Q},1}_2=\frac{e^2}{2m^2}\widetilde{\sum} \frac{\bra{\nu_R,\nu_L; 0}\B{A}_q\cdot\bm{\pi}\ket{\nu_R',\nu_L'; 1_{\B{k},\lambda}} \bra{\nu_R',\nu_L'; 1_{\B{k},\lambda}}\left[(\B{r}-\B{r}_0) \cdot \nabla \right]\sigma_z  {B}_{q,z}\ket{\nu_R,\nu_L; 0}}{-\omega + E_{\nu_L\vphantom{'},\nu_R\vphantom{'}}-E_{\nu_L',\nu_R'}} + \text{c.c.}\label{DeltaEQuad1}
\end{equation}
where $\bm{\sigma} \cdot \B{B}_q\to \sigma_z  {B}_{q,z}$ has been taken since the term in $\B{A}_q\cdot\bm{\pi}$ cannot change the spin state. Similarly, there is a quadrupole contribution from the application of the multipole operator to the term in $\B{A}_q\cdot\bm{\pi}$
\begin{equation}
\Delta E^{\text{Q},2}_2=\frac{e^2}{2m^2} \widetilde{\sum}\frac{\bra{\nu_R,\nu_L; 0}\left[(\B{r}-\B{r}_0) \cdot \nabla \right]\B{A}_q\cdot\bm{\pi}\ket{\nu_R',\nu_L'; 1_{\B{k},\lambda}} \bra{\nu_R',\nu_L'; 1_{\B{k},\lambda}}\sigma_z  {B}_{q,z}\ket{\nu_R,\nu_L; 0}}{-\omega + E_{\nu_L\vphantom{'},\nu_R\vphantom{'}}-E_{\nu_L',\nu_R'}} + \text{c.c.}\label{DeltaEQuad2}
\end{equation}
but which can contribute only for $\nu_R=\nu_R'$ and $\nu_L=\nu_L'$.
The other terms in Eqs.~\eqref{H2E} and \eqref{H2B} do not contribute because they result in terms whose order in the non-relativistic expansion, i.e.~in $1/m$, is higher than the leading $1/m^3$. Inserting the vector potential \eqref{AField} into Eqs.~\eqref{DeltaEQuad1} and \eqref{DeltaEQuad2} we find for the contributions proportional to $\sigma_z$
\begin{align}
\Delta \mathcal{E}^{\text{Q},1}_2&=\frac{e^2\sigma_z }{16m^2}\widetilde{\sum} \frac{h_i\Delta_i}{\Omega \omega} \; \frac{\Delta_i - (2\nu_i+1) \omega}{\omega^2-\Delta_i^2} \left( f_y\pd{^2f_y^*}{x^2}+  f_x\pd{^2f_x^*}{y^2}- f_x\pd{^2f_y^*}{x \partial y}-f_y\pd{^2f_x^*}{x \partial y} \right)  + \text{c.c.}\label{DeltaEQuad1F}\\
\Delta \mathcal{E}^{\text{Q},2}_2&= \frac{e^2}{2m^2}\sigma_z \widetilde{\sum}\left[h_i\nu_i+\frac{eB_0}{2m\Omega}\left(\nu_i+\frac{1}{2}\right)\right] \frac{|(\nabla \times \B{f})_z|^2}{2\omega^2}\;.\label{DeltaEQuad2F}
\end{align}
\end{widetext}

We now have the entire expression in terms of the mode functions $f_{\B{k}\lambda}$ of the part of the energy shift that is proportional to $\sigma_z$, i.e.~shifts the spin-up and down states differentially; it reads
\begin{align}
\Delta \mathcal{E}=&\; \Delta \mathcal{E}_1+\Delta \mathcal{E}_2^E+\Delta \mathcal{E}_2^B+\Delta \mathcal{E}^{\text{Q},1}_2+\Delta \mathcal{E}^{\text{Q},2}_2\label{EnergyShiftSum}
\end{align}
where the terms are given by Eqs.~\eqref{FirstOrderContribution}, \eqref{DeltaEDip}, \eqref{DeltaESpinTermsOfF}, \eqref{DeltaEQuad1F} and \eqref{DeltaEQuad2F}, respectively. 
It is tempting to simply expand Eq.~(\ref{DeltaEDip}) for a weak external magnetic field $B_0$ and extract a magnetic moment in the same way as our previous work \cite{BennettEberleinMagMomNJP, MagMomPRA}. However, we shall see in the following sections that this approach of calculating a magnetic moment is not necessarily physically appropriate for a bound electron. 

As an aside, we note that our calculation so far has included also the differential shift of the electron's spin-up and spin-down states that is independent of the magnetic field. 
While $\Delta\mathcal{E}_1 $ and $\Delta \mathcal{E}_2^B$ vanish for $B_0\rightarrow 0$, the sum
$\lim_{B_0\rightarrow 0}\left(\Delta \mathcal{E}_2^E+\Delta \mathcal{E}^{\text{Q},1}_2+\Delta \mathcal{E}^{\text{Q},2}_2\right)$ 
makes up the differential shift of the two spin states in the absence of any external magnetic field, which is easily written down from Eqs.~\eqref{DeltaEDip}, \eqref{DeltaEQuad1F}, and \eqref{DeltaEQuad2F}, and comes out proportional to $(\nu_R-\nu_L) \sigma_z e^2/m^2 $. The physics of this is easy to understand: this is just the $z$-dependent one-loop radiative self-energy correction $\Delta E^{\mbox{fs}}$ to the ${\bf L}\cdot{\bf S}$ fine-structure term $E^{\mbox{fs}}$ for an electron confined in a harmonic potential \cite{fn3},
$$
E^{\mbox{fs}}= (\nu_R-\nu_L) \sigma_z \frac{\omega_H^2}{4m}\;.
$$
However, in the following we shall only be interested in pure Zeeman shifts, i.e.~those depending on the magnetic field strength $B_0$.

\section{Relation to previous work}

In our previous work \cite{MagMomPRA} we considered an electron close to a surface but without any confining potential ($V_H=0$) and 
calculated the magnetic moment shift $\Delta \mu$ defined by $ \Delta \mathcal{E}_\text{free} = -\Delta \mu\; \sigma_z B_0$ in the limit of a weak external field $B_0\to0$. 
The result for $ \Delta \mathcal{E}_\text{free}$ in terms of mode functions was \cite{MagMomPRA,fn1}
\begin{widetext}
\begin{align}
 \Delta \mathcal{E}_\text{free}=\frac{e^3\sigma_z B_0}{4m^3}\SumOverModes\frac{1}{2\omega}\Bigg\{{ |f_z|^2}&+\frac{|{(\nabla \times \mathbf{f})}_x|^2}{\omega^2} + \frac{|{(\nabla \times \mathbf{f})}_y|^2}{\omega^2} \notag  \\
 &+\frac{1}{\omega^2}\left(f_x\frac{\partial^2f_y^* }{\partial x \partial y}+ f_y\frac{\partial^2f_x^*}{\partial x\partial y} -f_y\frac{\partial^2f_y^*}{\partial x^2}  -f_x\frac{\partial^2f_x^*}{\partial y^2}     + \text{c.c.} \right)\Bigg\} \label{ShiftInTermsofModeFunctions}
\end{align}
\end{widetext}
We can check the consistency of our method by looking at the small-trap frequency limit, $\omega_H\rightarrow 0$, of our results \eqref{EnergyShiftSum} in the presence of a harmonic confining potential $V_H$, Eq.~(\ref{VH}), and then looking for terms linear in the external field $B_0$, which should reproduce Eq.~\eqref{ShiftInTermsofModeFunctions}. We note that the limit $\omega_H\rightarrow 0$ implies
\begin{equation}
\Omega \to -\frac{eB_0}{2m}\;, \qquad \Delta_R \to -\frac{eB_0}{m}\;, \qquad \Delta_L \to 0
\label{omegatrapzero}
\end{equation}
in accordance with Eq.~(\ref{HHnotrap}).

We first consider $\Delta \mathcal{E}_2^E$, given by Eq.~\eqref{DeltaEDip} and take 
a small $\omegatrap$ expansion of it, finding
\begin{align}
\Delta \mathcal{E}_2^E=&\frac{e^3}{8m^2}B_0\sigma_z  \int d^3\B{k} \sum_{\lambda}\left(|f_x|^2+|f_y|^2\right)\notag \\
& \times \frac{ eB_0 (2  \nu_R+1)+m \omega }{ e^2 B_0^2  - m^2\omega ^2}+\mathcal{O}(\omegatrap^2)
\end{align}
Then taking a small $B_0$ expansion of the leading term in this series, we find for the term linear in $B_0$
\begin{align}
\Delta \mathcal{E}_{2,\text{free}}^E=-\frac{e^3}{4m^3}B_0\sigma_z  \SumOverModes \frac{|f_x|^2+|f_y|^2}{2\omega} \label{ExpansionIntermediateStep}
\end{align}
where the subscript `free' indicates that this is an expression in the limit of vanishing confinement potential $V_H$ and which should facilitate a consistency check with Eq.~\eqref{ShiftInTermsofModeFunctions}. Combining this with the first-order contribution in Eq.~\eqref{FirstOrderContribution} (which is independent of $\omegatrap$, and already linear in $B_0$, so that $\Delta \mathcal{E}_{1,\text{free}}= \Delta \mathcal{E}_{1}$), we find
\begin{align}
 \Delta \mathcal{E}_{1,\text{free}}+\Delta \mathcal{E}_{2,\text{free}}^E=\frac{e^3}{4m^3}B_0\sigma_z  \SumOverModes \frac{|f_z|^2}{2\omega} \label{DeltaE2ESmallB}
\end{align}
in agreement with the first term in Eq.~\eqref{ShiftInTermsofModeFunctions}. 

Having considered the first two terms of Eq.~\eqref{EnergyShiftSum}, we now move on to the third term $\Delta \mathcal{E}^B_2$ given by Eq.~\eqref{DeltaESpinTermsOfF}. This is independent of $\omegatrap$ and already linear in $B_0$ so that
\begin{equation}
\Delta \mathcal{E}_{2,\text{free}}^B \equiv \Delta \mathcal{E}_{2}^B
\label{DeltaESpinSmallB}
\end{equation}
and without further manipulation this gives the second and third terms of Eq.~\eqref{ShiftInTermsofModeFunctions}.

Next we move on to the quadrupole terms given by Eqs.~\eqref{DeltaEQuad1F} and \eqref{DeltaEQuad2F}. Proceeding in an identical way to that which produced Eq.~\eqref{ExpansionIntermediateStep}, we have for the term linear in $B_0$ in the small $\omegatrap$ approximation of Eq.~\eqref{DeltaEQuad1F}
\begin{widetext}
\begin{equation}
\Delta \mathcal{E}^{\text{Q},1}_{2,\text{free}}=-\frac{e^3\sigma_zB_0 }{8m^3}\SumOverModes \frac{1}{\omega^3} \left( f_y\pd{^2f_y^*}{x^2}+  f_x\pd{^2f_x^*}{y^2}- f_x\pd{^2f_y^*}{x \partial y}-f_y\pd{^2f_x^*}{x \partial y} \right)  + \text{c.c.} \label{DeltaEQuadSmallB}
\end{equation}
\end{widetext}
which agrees with the final term in Eq.~\eqref{ShiftInTermsofModeFunctions}. 

Finally, $\Delta \mathcal{E}^{\text{Q},2}_2$ of Eq.~\eqref{DeltaEQuad2F} does not contribute since its $\omegatrap\to 0$ limit is independent of the magnetic field $B_0$, so that \cite{fn2}
\begin{equation}
\Delta \mathcal{E}^{\text{Q},2}_{2, \text{free}}=0 \label{DeltaEQuad2FZero}
\end{equation}
Combining Eqs.~\eqref{DeltaE2ESmallB}, \eqref{DeltaESpinSmallB}, \eqref{DeltaESpinTermsOfF}, \eqref{DeltaEQuadSmallB} and \eqref{DeltaEQuad2FZero} we see that the sum
\begin{align}
\Delta \mathcal{E}_{1,\text{free}}+\Delta \mathcal{E}_{2,\text{free}}^E+\Delta \mathcal{E}_{2,\text{free}}^B+\Delta \mathcal{E}^{\text{Q},1}_{2,\text{free}}+\Delta \mathcal{E}^{\text{Q},2}_{2,\text{free}}\end{align}
does indeed reproduce Eq.~\eqref{ShiftInTermsofModeFunctions}, i.e.~in a weak magnetic field the term linear in $B_0$ of the vanishing-trap limit $\omegatrap\to 0$ of Eq.~\eqref{EnergyShiftSum} agrees with magnetic-moment shift calculated in Refs.~\cite{MagMomPRA}.

Therefore, we have shown that the energy shift \eqref{ShiftInTermsofModeFunctions} of an electron not subject to harmonic confinement can be reproduced as a special case of the energy shift of the confined electron, Eq.~\eqref{EnergyShiftSum}, which represents an important consistency check on our results.
 
\section{Evaluation of the energy shift}
For the purposes of evaluating and analysing the shift, we split the total energy shift \eqref{EnergyShiftSum} into two distinct contributions,
\begin{equation}
\Delta \mathcal{E} = \Delta \mathcal{E}_D+\Delta \mathcal{E}_S \label{DSDefinition}
\end{equation}
with
\begin{align}
\Delta \mathcal{E}_D=&\; \Delta  \mathcal{E}_2^E +\Delta \mathcal{E}_2^{\text{Q},1} \label{DeltaEL} \\
\Delta \mathcal{E}_S=&\;\Delta \mathcal{E}_1+ \Delta \mathcal{E}_2^B+\Delta \mathcal{E}_2^{\text{Q},2}\;. \label{DeltaES}
\end{align}
The first part, $\Delta \mathcal{E}_D$, contains the terms that contribute at $\{\nu_L',\nu_R'\}\neq \{\nu_L,\nu_R\}$ (i.e. which have arisen from virtual transitions between Landau levels), and the second part, $\Delta \mathcal{E}_S$, contains the terms that contribute at $\{\nu_L',\nu_R'\}=\{\nu_L,\nu_R\}$ (i.e. which come from virtual spin flips within the same Landau level).

None of our discussions so far have been in any way specific to the shape or character of the surface near which the electron is trapped; Eqs.~\eqref{FirstOrderContribution}, \eqref{DeltaEDip}, \eqref{DeltaESpinTermsOfF}, \eqref{DeltaEQuad1F} and \eqref{DeltaEQuad2F} for the various parts entering Eqs.~(\ref{DeltaEL}) and (\ref{DeltaES}) are valid for any quantized field coupled to an electron confined by $V_H$ and subject to the external magnetic field $B_0$. From now on, as an example, we shall consider a surface filling the space $z>0$, as shown in Figure \ref{HalfSpace} and assume the response of the surface to electromagnetic radiation to be described by its dielectric permittivity $\epsilon(\omega)$. We shall initially consider the surface to be non-dispersive, which means that it is described by a single number $n$, its refractive index, defined through $\epsilon(\omega) = n^2$. This model has the advantage of allowing the quantized field near such a surface to be written down in terms of an explicit mode expansion, as detailed in Appendix \ref{ModesAppendix}.

\subsection{Shift due to transitions between different Landau levels}\label{DiffLandauLevelsSection}

\begin{figure}
\vspace*{-40mm}
\includegraphics[width = 0.45\textwidth]{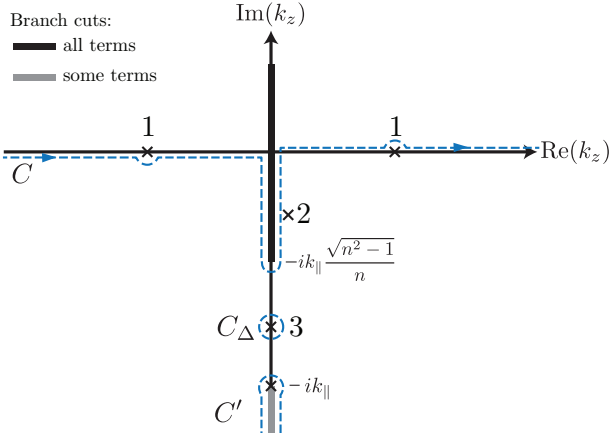}
\caption{\label{ContourDielectricHBE} Lower complex $k_z$ plane for the integrals in Eqs.~\eqref{DeltaEDipRenormExplicit} and \eqref{DeltaEQuadRenormExplicit}. The poles at $k_z^2 = \Delta_i^2 - k_\parallel^2$ can appear in one of three positions depending on the relative values of $k_\parallel$, $\Delta_i$ and $n\Delta_i$. These positions are at equal and opposite points on the real axis (shown as positions 1), and on the positive and negative imaginary axis, either to the side of the cut due to $k_z^d$ (2), or between the two cuts (3). Poles in the upper half-plane are not shown as they are irrelevant for the calculation at hand.}
\end{figure}

To evaluate $\Delta \mathcal{E}_D$ initially for a non-dispersive dielectric, we substitute the non-dispersive modes, Eqs.~\eqref{NonDispModes}, into Eqs.~\eqref{DeltaEDip} and \eqref{DeltaEQuad1F}, and obtain
\begin{widetext}
\begin{align}
\Delta  \mathcal{E}_2^E =& -\frac{1}{(2\pi)^3}\frac{e^2}{8m^2\Omega}\sigma_z \sum_{\lambda,i,\vartheta} h_i \Delta_i^2 \int d^2\mathbf{k}_\parallel \Bigg\{   \int_0^\infty dk_z \, \alpha^\vartheta_\lambda[1+|R^\text{vac}_\lambda|^2]+\frac{1}{n^2} \int_{-\infty}^{-\sqrt{n^2-1}k_\|} dk_z^d \alpha^\vartheta_\lambda|T^\text{med}_\lambda|^2  \notag \\
&\qquad+ \vartheta\int_0^\infty dk_z \, \alpha^\vartheta_\lambda  R^\text{vac}_\lambda(e^{2ik_zz}+e^{-2ik_zz})+ \frac{\vartheta}{n^2} \int_{-\sqrt{n^2-1}k_\|}^0 dk_z^d \alpha^\vartheta_\lambda|T^\text{med}_\lambda|^2 e^{2ik_z z} \Bigg\}  \frac{\Delta_i(2\nu_i+1)-\omega}{\omega^2-\Delta_i^2}  \label{DeltaEDipForm}\\
\Delta \mathcal{E}_2^{\text{Q},1} =&-\frac{1}{(2\pi)^3}\frac{e^2}{8m^2\Omega}\sigma_z \sum_{\lambda,i,\vartheta} h_i\Delta_i  \int d^2\mathbf{k}_\parallel \Bigg\{   \int_0^\infty dk_z \, \beta^\vartheta_\lambda[1+|R^\text{vac}_\lambda|^2]+\frac{1}{n^2} \int_{-\infty}^{-\sqrt{n^2-1}k_\|} dk_z^d \beta^\vartheta_\lambda|T^\text{med}_\lambda|^2 \notag \\
&\qquad + \vartheta\int_0^\infty dk_z \, \beta^\vartheta_\lambda  R^\text{vac}_\lambda(e^{2ik_zz}+e^{-2ik_zz})+ \frac{\vartheta}{n^2} \int_{-\sqrt{n^2-1}k_\|}^0 dk_z^d \beta^\vartheta_\lambda|T^\text{med}_\lambda|^2 e^{2ik_z z} \Bigg\}\frac{\Delta_i - (2\nu_i+1) \omega}{\omega(\omega^2-\Delta_i^2)} \label{DeltaEQuadForm} \end{align}
\end{widetext}
where the summation is over polarization $\lambda$, handedness $i$ and a new index $\vartheta=\pm1$, introduced to make the expressions in Eqs.~\eqref{DeltaEDipForm} and \eqref{DeltaEQuadForm} less cumbersome. The sum stands for
\begin{equation}
\sum_{\lambda,i,\vartheta}  \equiv 
    \sum_{{\substack{\lambda = \text{TE,TM}\\
                    {i = L,R}\\
                  {\vartheta = -1,+1}}}} 
\end{equation}
and the coefficients $\alpha^\vartheta_\lambda$ and $\beta^\vartheta_\lambda$ are
\begin{align}
\alpha_{\text{TE}}^+  &= \frac{1}{2}, & \alpha_{\text{TM}}^-  &=\frac{k_z^2}{2k^2}, &\beta_{\text{TE}}^+  &= k_\parallel^2,\notag \\
&&&&& \hspace*{-54mm} {\{ \alpha_{\text{TE}}^-,\alpha_{\text{TM}}^+ ,\beta_{\text{TE}}^- ,\beta_{\text{TM}}^+ ,\beta_{\text{TM}}^- \}=0}\label{AlphaBetaDefs}\end{align}
Following \cite{PRDEberleinRobaschik,BennettEberleinMagMomNJP,MassShiftsPaper,MagMomPRA}, we use the relation ${dk_z^d}=n^2 (k_z/k_z^d) dk_z$ to manipulate the $k_z$ integral in the first line of Eq.~\eqref{DeltaEDipForm} to 
\begin{align}
\int_0^\infty dk_z \, \alpha^\vartheta_\lambda &\left[1+|R^\text{vac}_\lambda|^2 + \frac{k_z}{k_z^d}  |T^\text{med}_\lambda|^2 \right] \notag \\
&= 2\int_0^\infty dk_z \, \alpha^\vartheta_\lambda\;, \label{RTRelation}
\end{align}
where the equality follows since $k_z$ and $k_z^d$ are here both real and represents current conservation for modes with only travelling waves.
The $k_z$ integral in the second line of Eq.~(\ref{DeltaEDipForm}) is $z$ dependent as it arises from the interference of incident and reflected waves; it can be written as:
\begin{align}
 \vartheta\int_0^\infty dk_z \, \alpha^\vartheta_\lambda &  R^\text{vac}_\lambda(e^{2ik_zz}+e^{-2ik_zz}) \notag \\
  + &\vartheta \int_{0}^{i\sqrt{n^2-1}k_\|/n} dk_z \frac{k_z}{k_z^d} \alpha^\vartheta_\lambda |T^\text{med}_\lambda|^2 e^{2ik_z z}\;. \label{TwoIntegrals}
\end{align}
We observe that for real $k_z^d$ and pure imaginary $k_z$ the following relation holds for either polarization $\lambda$,
\begin{equation}
R^\text{vac}_{\lambda} |_{ k_z^d = -K} - R^\text{vac}_{\lambda}|_{k_z^d = K} = \frac{k_z}{k_z^d} T^\text{med}_\lambda T^{\text{med}*}_\lambda |_{ k_z^d = -K}\;,
\label{relation}
\end{equation}
which permits us to combine the integrals in Eq.~(\ref{TwoIntegrals}) into one,
\begin{equation}
\vartheta \int_C dk_z \,  \alpha^\vartheta_\lambda R^\text{vac}_\lambda e^{2ik_z z}\;.
\end{equation}
with the contour $C$ as shown in Fig.~\ref{ContourDielectricHBE}. Rearranging Eq.~\eqref{DeltaEQuadForm} in precisely the same way, we arrive at
\begin{widetext}
\begin{align}
\Delta  \mathcal{E}_2^E  =& -\frac{1}{(2\pi)^3}\frac{e^2}{8m^2\Omega}\sigma_z \sum_{\lambda,i,\vartheta}h_i \Delta_i^2\int d^2\mathbf{k}_\parallel \Bigg\{  \vartheta \int_C dk_z \,  \alpha^\vartheta_\lambda R^\text{vac}_\lambda e^{2ik_z z}+2\int_0^\infty dk_z \, \alpha^\vartheta_\lambda\Bigg\}   \frac{\Delta_i(2\nu_i+1)-\omega}{\omega^2-\Delta_i^2} \label{DeltaEDipSimp}\\
\Delta \mathcal{E}_2^{Q,1}=&-\frac{1}{(2\pi)^3}\frac{e^2}{8m^2\Omega}\sigma_z\sum_{\lambda,i,\vartheta} {h_i\Delta_i}\int d^2\mathbf{k}_\parallel \Bigg\{  \vartheta \int_C dk_z \,  \beta^\vartheta_\lambda R^\text{vac}_\lambda e^{2ik_z z}+2\int_0^\infty dk_z \, \beta^\vartheta_\lambda\Bigg\} 
\frac{\Delta_i - (2\nu_i+1) \omega}{\omega(\omega^2-\Delta_i^2)}\label{DeltaEQuadSimp} \end{align}
\end{widetext}
with the integration path $C$ as shown in Fig.~\ref{ContourDielectricHBE}.
The second terms in the brackets in Eqs.~\eqref{DeltaEDipSimp} and \eqref{DeltaEQuadSimp} are independent of $z$, i.e.~they would be present even in the absence of the surface. These are free-space counter\-terms that we need to subtract, since we are interested only in the surface-dependent Zeeman shift of the spin energy levels. Subtracting them and at the same time substituting the explicit expressions for the various coefficients from Eq.~\eqref{AlphaBetaDefs}, we obtain for the renormalized position-dependent energy shifts
%
%
\begin{align}
\Delta  \mathcal{E}_2^E &= -\frac{e^2}{8m^2\Omega}\frac{1}{8\pi^2}\sigma_z \sum_{i}h_i\Delta_i^2\int_0^\infty d{k}_\parallel k_\parallel \int_C dk_z   \notag \\
&\times   \frac{\Delta_i(2\nu_i+1)-\omega}{\omega^2-\Delta_i^2}\left( R^\text{vac}_\text{TE}-\frac{k_z^2}{\omega^2 }R^\text{vac}_\text{TM}  \right)  e^{2ik_z z} \label{DeltaEDipRenormExplicit}\\
\Delta \mathcal{E}_2^{Q,1}&=-\frac{e^2}{8m^2\Omega}\frac{1}{4\pi^2}\sigma_z \sum_{i}h_i\Delta_i\int_0^\infty d{k}_\parallel \, k_\parallel^3  \int_C dk_z \notag \\
&  \times   \frac{\Delta_i - (2\nu_i+1) \omega}{\omega(\omega^2-\Delta_i^2)}R^\text{vac}_\text{TE} e^{2ik_z z}
\label{DeltaEQuadRenormExplicit} 
\end{align}
The structure of the complex $k_z$ plane for the integrals in Eqs.~\eqref{DeltaEDipRenormExplicit} and \eqref{DeltaEQuadRenormExplicit} is shown in Fig.~(\ref{ContourDielectricHBE}). 
There is a branch cut due to $k_z^d= \sqrt{n^2(k_z^2+k_\parallel^2)-k_\parallel^2}$, which we have placed between the two branching points at $k_z=\pm ik_\parallel\frac{\sqrt{n^2-1}}{n}$ in order to make use of relation (\ref{relation}) that allows us to combine the contribution from evanescent modes into one integral with that from travelling modes if we integrate along the path $C$ \cite{PRDEberleinRobaschik}. There is also a branch cut due to $\omega=\sqrt{k_z^2+k_\parallel^2}$ in the denominators of all but one of the terms in Eqs.~\eqref{DeltaEDipRenormExplicit} and \eqref{DeltaEQuadRenormExplicit}, which we place along $k_z = \pm ik_\parallel... \pm i \infty $. 

Furthermore, there are two poles at $k_z=\pm\sqrt{\Delta_i^2 - k_\parallel^2}$ whose positions move through three distinct regions as $k_\parallel$ is integrated over. 
For $k_\parallel<\Delta_i$ they lie at equal and opposite points on the real axis (position 1 in Fig.~\ref{ContourDielectricHBE}), for $\Delta_i<k_\parallel< n\Delta_i$ they appear on the positive and negative imaginary axes to the side of the cut due to $k_z^d$ (position 2 in Fig.~\ref{ContourDielectricHBE}), and for $k_\parallel> n \Delta_i$ they lie between $\pm ik_\parallel\frac{\sqrt{n^2-1}}{n}$ and $\pm ik_\parallel$ (position 3 in Fig.~\ref{ContourDielectricHBE}). 

In order to understand how the integration path $C$ circumvents these poles, one first needs to understand the physical nature of these poles. Any of the excited Landau states ($\nu_R>0$ and/or $\nu_L>0$) can decay to the ground state ($\nu_R=0=\nu_L$), so that these transitions are not just virtual in second-order perturbation theory but real, and thus they give rise to poles in the sum in Eq.~(\ref{SecondOrderPertOnly}). For these excited states the quantity $\Delta E$ is complex; its real part is the energy shift and its imaginary part gives the decay rate. Here we are interested only in the energy shift, which is obtained either by taking the principal value of the $k_z$ integral around the poles at $k_z^2 = \Delta_i^2 - k_\parallel^2$ or by choosing the integration path $C$ around the poles as shown in Fig.~\ref{ContourDielectricHBE} and taking the real part of the integral. For the poles in position 1 (see Fig.~\ref{ContourDielectricHBE}) one needs to choose the path $C$ to run above one of the poles and below the other; and for position 2 the path $C$ needs to run up and down either both to the left of the pole or both to the right of the pole in the lower half-plane (cf.~Fig.~\ref{ContourDielectricHBE}). Then it is straightforward to show that the sum of the residue contributions around the pole(s) is pure imaginary in all cases, so that calculating the integral along path $C$ and then taking the real part is the same as taking its principal value. Of course, for $\nu_i=0$ the pole disappears because the term $\omega-\Delta_i$ cancels between numerators and denominators in Eqs.~\eqref{DeltaEDipRenormExplicit} and \eqref{DeltaEQuadRenormExplicit}, which is what one expects as the ground state cannot decay.

We now proceed to deform the contour $C$ into the lower half-plane, picking up two contributions, one along the path $C_\Delta$ around the pole, which, depending on $k_\|$, can be either the right pole in position 1, or the lower half-plane ones in positions 2 and 3, and the second along the path $C'$ around the branch cut due to $\omega=\sqrt{k_z^2+k_\|^2}$, as shown in Fig.~\ref{ContourDielectricHBE}. 
\begin{align}
\Delta  \mathcal{E}_2^E =& -\frac{e^2}{8m^2\Omega}\frac{1}{8\pi^2}\sigma_z \sum_{i} h_i\Delta_i^2 \int_0^\infty d{k}_\parallel k_\parallel\notag \\
&\times  \left(\int_{C_\Delta}dk_z  +\int_{C'} dk_z  \right)  \frac{\Delta_i(2\nu_i+1)-\omega}{\omega^2-\Delta_i^2} \notag \\
&\times  \left( R^\text{vac}_\text{TE}-\frac{k_z^2}{\omega^2 }R^\text{vac}_\text{TM}  \right)  e^{2ik_z z} \label{DeltaEDipRenormExplicitDeformed}\\
\Delta \mathcal{E}_2^{Q,1}=&-\frac{e^2}{8m^2\Omega}\frac{1}{4\pi^2}\sigma_z \sum_{i} h_i \Delta_i\int_0^\infty d{k}_\parallel \, k_\parallel^3 \notag \\
&\times   \left(\int_{C_\Delta}dk_z  +\int_{C'} dk_z  \right) \frac{\Delta_i - (2\nu_i+1) \omega}{\omega(\omega^2-\Delta_i^2)}\notag \\
&  \times  R^\text{vac}_\text{TE}\; e^{2ik_z z}\label{DeltaEQuadRenormExplicitDeformed} \end{align}
Both these terms together give the energy shift $\Delta \mathcal{E}_D$, as defined by Eq.~\eqref{DeltaEL}, due to transitions between Landau levels. 

\subsection{Shift due to transitions within the same Landau level}\label{Sec:Qzero}

Proceeding along the same lines that led to Eqs.~\eqref{DeltaEDipRenormExplicit} and \eqref{DeltaEQuadRenormExplicit} in the previous section, we find for $\Delta \mathcal{E}_2^{Q,2}$ from Eq.~\eqref{DeltaEQuad2F}
\begin{align*}
\Delta \mathcal{E}_2^{Q,2}=&\frac{e^2}{4m^2}\frac{1}{4 \pi^2}\sigma_z \sum_i \left[h_i\nu_i+\frac{eB_0}{2m\Omega}\left(\nu_i+\frac{1}{2}\right)\right]  \notag \\
&\times  \int_0^\infty d{k}_\parallel \, k_\parallel^3 \int_{C}dk_z\; \frac{R^\text{vac}_\text{TE}}{\omega^2}\; e^{2ik_z z}
\end{align*}
Since the TE reflection coefficient vanishes at $k_z=-ik_\parallel$, the integrand is analytic in the lower complex $k_z$ plane, and hence $\Delta \mathcal{E}_2^{Q,2}$ is zero. This makes sense physically, because Eq.~(\ref{DeltaEQuad2}) shows that for this term the virtual photon transition does not couple to any transition between Landau levels or spin flip.

Repeating the analysis of Sec.~\ref{DiffLandauLevelsSection} for the remaining terms of $\Delta \mathcal{E}_S$, we find 
\begin{align}
\Delta \mathcal{E}_1=&\frac{e^3B_0}{8m^3 }\frac{1}{4 \pi^2}\sigma_z  \int_0^\infty dk_\parallel \; k_\parallel \int_{C'} dk_z \notag \\
&\times \frac{1}{\omega}\left[ R^\text{vac}_{\text{TE}} + \frac{1}{\omega^2}\left( k_\parallel^2-{k_z^2}\right)R^\text{vac}_{\text{TM}}  \right]e^{2ik_z z} \label{DeltaEFirstOrderFinal}\\
\Delta \mathcal{E}_2^B =& \frac{e^3B_0}{8m^3}\frac{1}{4 \pi^2} \sigma_z\int_0^\infty dk_\parallel \; k_\parallel \int_{C'} dk_z  \notag \\
&\times \frac{1}{\omega^3}\left[-R^\text{vac}_{\text{TE}}k_z^2 + (k_z^2+k_\parallel^2)R^\text{vac}_{\text{TM}} \right] e^{2i k_z z}\label{DeltaESpinFinal} \end{align}
where the contour $C'$ is as shown in Fig.~\ref{ContourDielectricHBE}.

\subsection{Total}
The sum of Eqs.~\eqref{DeltaEDipRenormExplicitDeformed}--\eqref{DeltaESpinFinal} gives the energy shift of the two spin states. Before discussing the details of this result for a perfect reflector and then for a non-dispersive dielectric, we note that we have previously shown \cite{MagMomPRA} that, in the absence of any confining potential, derivations that were specific to a non-dispersive dielectric nevertheless result in formulae for energy level shifts in terms of integrals over reflection coefficients that are in fact valid for dispersive dielectrics, crudely speaking because reflection coefficients have some quite general properties. This also can be shown explicitly by employing a noise-current approach \cite{thesis}. The same arguments apply to the present calculation; the formulae in Eqs.~\eqref{DeltaEDipRenormExplicitDeformed}--\eqref{DeltaESpinFinal} are applicable not just to the specific model detailed in Appendix \ref{ModesAppendix} but also to dispersive dielectrics, i.e.~one
can simply use the same formulae but replace the reflection coefficients with dispersive ones. 
However, the overwhelming complexity of the integrals for a dispersive medium means that in the following we shall limit ourselves to presentation of results for a perfect reflector and for a non-dispersive dielectric since these are accessible analytically and demonstrate the main features of the model. But, we emphasize that Eqs.~\eqref{DeltaEDipRenormExplicitDeformed}--\eqref{DeltaESpinFinal} are also valid for dispersive dielectrics, as explained in detail in \cite{MagMomPRA}. 

\section{Zeeman shift near a perfect reflector}\label{sec:perfectrefl}
The simplest model of a reflecting surface is a perfectly reflecting plane. The mode functions of the electromagnetic field in its presence can be obtained from those given in Appendix \ref{ModesAppendix}, Eq.~(\ref{NonDispModes}), by taking the limit $n\rightarrow\infty$, which shows instantly that for a perfect reflector there are no modes incident from the side of the medium. The reflection coefficients turn into:
\begin{align}
\lim_{n\rightarrow\infty} R^\text{vac}_{\text{TE}} = -1\ , \quad
\lim_{n\rightarrow\infty} R^\text{vac}_{\text{TM}} = 1\ .
\label{Rlimits}
\end{align}
Substituting these limits into Eqs.~\eqref{DeltaEDipRenormExplicitDeformed}--\eqref{DeltaESpinFinal}, we find that the integrals simplify significantly and we can carry out all the integrations analytically. To calculate the contribution from the path $C_\Delta$ to $\Delta \mathcal{E}_2^{E}$ and $\Delta \mathcal{E}_2^{Q,1}$ in Eqs.~\eqref{DeltaEDipRenormExplicitDeformed} and \eqref{DeltaEQuadRenormExplicitDeformed}, we calculate the residue, which is straightforward, and then carry out the remaining integration over $k_\parallel$ by changing variables from $k_\parallel$ to
\begin{align}
x= \frac{\sqrt{\Delta_i^2-k_\parallel^2}}{\Delta_i} \quad \mbox{and}\ \quad y= \frac{\sqrt{k_\parallel^2-\Delta_i^2}}{\Delta_i} 
\label{substitutions}
\end{align}
for $k_\parallel <\Delta_i$ and $k_\parallel >\Delta_i$, respectively. Taking the real part of the result, we obtain
\begin{align}
\Delta \mathcal{E}_D(C_\Delta)=& \Delta \mathcal{E}_2^E(C_\Delta) + \Delta \mathcal{E}_2^{Q,1}(C_\Delta)\notag\\
=& \frac{e^2\sigma_z}{128\pi m^2\Omega z^3} \sum_i h_i \Delta_i\nu_i \big[ 6z\Delta_i \sin 2z\Delta_i \notag\\
&+(3-4z^2\Delta_i^2) \cos 2z\Delta_i \big]\;.
\label{perfectCdelta}
\end{align}
To calculate the contribution from the path $C'$, we note that only terms that are odd in $\omega$ contribute, as there is no square root cut for those that are even in $\omega$, and that 
$\omega=-i\sqrt{-k_z^2-k_\|^2}$ to the right of the cut in the lower half $k_z$ plane and $\omega=+i\sqrt{-k_z^2-k_\|^2}$ to the left of it. Renaming $k_z=-i\kappa$ we find for the contribution from $C'$ to $\Delta \mathcal{E}_2^{Q,1}$ 
\begin{align}
\Delta \mathcal{E}_2^{Q,1}(C') =& - \frac{e^2\sigma_z}{16\pi^2 m^2\Omega} \sum_i h_i \Delta_i^2 \int_0^\infty dk_\parallel \; k_\parallel^3\notag\\
&\times \int_{k_\|}^\infty d\kappa\; \frac{\sqrt{\kappa^2-k_\|^2}}{(k_\|^2-\kappa^2)(k_\|^2-\kappa^2-\Delta_i^2)}\; e^{2\kappa z} \notag
\end{align}
Upon interchanging the order of integrations 
$$
\int_0^\infty dk_\parallel \int_{k_\|}^\infty d\kappa \longrightarrow \int_0^\infty d\kappa \int_0^\kappa dk_\parallel
$$
and then changing variables from $k_\|$ to $\xi=\sqrt{\kappa^2-k_\|^2}$, the $\xi$ integral can be carried out to give powers of $\kappa$ and an $\arctan(\kappa/\Delta_i)$. Then the $\kappa$ integral can be carried out, giving combinations of sine and cosine integrals \cite{Abramowitz} with sines and cosines. Doing the same also for $\Delta \mathcal{E}_2^{E}$, we find for the total contributions from the path $C'$
\begin{align}
\Delta & \mathcal{E}_D(C')= \Delta \mathcal{E}_2^E(C') + \Delta \mathcal{E}_2^{Q,1}(C')\notag\\
=& -\frac{e^2\sigma_z}{128\pi^2 m^2\Omega z^3} \sum_i h_i \Delta_i \Big\{ 2z\Delta_i \notag\\
&+ \mbox{si}(-2z\Delta_i) \big[ 6z\Delta_i \sin 2z\Delta_i +(3-4z^2\Delta_i^2) \cos 2z\Delta_i \big]\notag\\
&+ \mbox{Ci}(-2z\Delta_i) \big[  (3-4z^2\Delta_i^2) \sin 2z\Delta_i - 6z\Delta_i \cos 2z\Delta_i\big]\Big\}\;.
\label{perfectCprime}
\end{align}

Finally we work out the energy shift due to spin flips, we substitute the limits in Eq.~\eqref{Rlimits} into Eqs.~\eqref{DeltaEFirstOrderFinal} and \eqref{DeltaESpinFinal}, which gives
\begin{equation*}
\Delta \mathcal{E}_1 + \Delta \mathcal{E}_2^B =\frac{e^3B_0\sigma_z }{32 \pi^2 m^3 }  \int_0^\infty dk_\parallel \; k_\parallel^3 \int_{C'} dk_z \frac{e^{2ik_z z}}{\omega^3}
\end{equation*}
Since in this limit the integrand is not dependent on $k_z^d$ any longer, there is no square root cut due to $k_z^d$ and we can deform the contour $C'$ back to run along the real $k_z$ axis. Carrying out the $k_z$ integration then gives a Bessel function $\text{K}_1(-2k_\parallel z)$ and the subsequent $k_\parallel$ integration yields
\begin{equation}
\Delta \mathcal{E}_1 + \Delta \mathcal{E}_2^B =\frac{e^3B_0\sigma_z }{32 \pi^2 m^3 z^2}\;. 
\label{perfectSpinflip}
\end{equation}

Therefore the total Zeeman energy shift of a trapped electron near a perfectly reflecting plane is given by the sum of Eqs.~\eqref{perfectCdelta}, \eqref{perfectCprime}, and \eqref{perfectSpinflip},
\begin{align}
\Delta \mathcal{E} = -\frac{e^2\sigma_z}{128\pi^2 m^2 z^3} \left[ -4z\frac{eB_0}{m}  + \sum_i h_i \frac{\Delta_i}{\Omega}\; \mathcal{F}_i(-2z\Delta_i)\right]
\label{perfectShift}
\end{align}
with the dimensionless function, parametrically dependent on the quantum numbers $\nu_R$ and $\nu_L$ of the state,
\begin{align}
\mathcal{F}_i(\theta) = &-\theta + \mbox{Ci}(\theta)\left[ 3\theta \cos\theta -(3-\theta^2) \sin\theta \right]\notag\\
& + \left[\mbox{si}(\theta)-\pi\nu_i \right] \left[ (3-\theta^2)\cos\theta + 3\theta\sin\theta\right]\;.
\end{align}
The abbreviations $\Omega$, $\Delta_i$, and $h_i$ are defined in Eqs.~\eqref{OmegaDef}, \eqref{DeltaDef}, and \eqref{hDef}, respectively. The function $\mathcal{F}_i(\theta)$ is linear for small arguments
\begin{equation}
\mathcal{F}_i(\theta\ll 1) = -3\pi\left(\nu_i +\frac12\right) + 2\theta + O(\theta^2)
\label{smalltheta}
\end{equation}
and oscillates with a quadratically growing amplitude for large arguments
\begin{equation}
\mathcal{F}_i(\theta\gg 1) = \pi\nu_i \left( \theta^2\cos\theta -3 \theta\sin\theta -3\cos\theta \right) -\frac8\theta +O(\theta^{-3})
\label{largetheta}
\end{equation}
unless the particle is in the ground state of the trap, $\nu_i=0$, when $\mathcal{F}_i(\theta)$ falls as $\theta^{-1}$. Since $\theta=-2z\Delta_i$ in Eq.~\eqref{perfectShift} for the Zeeman energy shift, this means that at short distances from the wall, the shift is roughly the same in ground and excited states of the trap, but at large distances the shift of excited states is orders of magnitude bigger than that of the ground state.
We shall discuss this behaviour of the shift in more detail together with asymptotic expressions for the shift near a dielectric surface in the following section.

\section{Asymptotic regimes} \label{AsymptoticRegimesHBE}
 
For a free electron near a surface \cite{MagMomPRA} the weak-field or  `non-retarded' regime is in S.I. units defined by
\begin{equation}
\frac{|e| B_0}{m} \ll \frac{c}{|z|} \; , \label{NonRetardedDefinition}
\end{equation}
Typical magnetic field strengths used in experiments with trapped electrons are relatively strong, usually of the order of a few Tesla \cite{HaffnerMagMom, Werth2013}, giving
\begin{equation}
\frac{|e| B_0}{m} \sim 10^{11} \text{Hz} \; . 
\end{equation}
The weak-field regime according to Eq.~\eqref{NonRetardedDefinition} then applies to $|z|\ll 3$mm, which is comfortably within the reach of modern trapping technology. For an electron in a trap the  trap frequency $\omegatrap$ provides an additional scale, leading to three different kinds of `weak-field' regimes to be discussed below. 

With regard possible values of $\omegatrap$ in applications, we shall concentrate on two realistic settings. The first of these is an electron in a Penning trap, for which the closest analogue of our trap frequency is the magnetron oscillation frequency, which is of order $100$kHz (see, for example, \cite{GabrielsegfactorPRL}). The second is an electron bound in an atom; for a hydrogen atom the frequency of the `trap' is around a few eV, corresponding to $\omegatrap\sim10^{15}$Hz. 

Setting the trap frequency in relation to the other two scales, the cyclotron frequency and the inverse of the distance from the surface, we have to distinguish three cases in the weak-field regime.
\subsection{Small trap frequency: $\omegatrap \ll \frac{|e| B_0}{m}  \ll \frac{c}{|z|}$}
The constraint $ \omegatrap \ll \frac{|e| B_0}{m} $ means that Penning traps are the most relevant type of binding potential, based on the numerical values given above. Since $\omegatrap$ is  small compared to all other scales, the trapping potential is very weak and one expects no significantly new behaviour relative to the free-space case, analysed in detail in Ref.~\cite{MagMomPRA}. 
Indeed, taking the limit $\omegatrap\rightarrow 0$ in Eq.~\eqref{perfectShift} implies the limits listed in Eq.~\eqref{omegatrapzero}, and therefore $\frac{|e| B_0}{m}  \ll \frac{c}{|z|}$ implies $|z|\Delta_i\ll 1$, so that the small-argument asymptotics given in Eq.~\eqref{smalltheta} applies. The first term in Eq.~\eqref{smalltheta} gives just a $B_0$ independent term, but the second term in Eq.~\eqref{smalltheta} together with the rest of Eq.~\eqref{perfectShift} reproduces the magnetic moment shift of a free particle near a perfectly reflecting surface in Eq.~(B1) of Ref.~\cite{MagMomPRA} and derived earlier in Ref.~\citep{BartonFawcett}. Additionally, Ref.~\cite{MagMomPRA} provides a comprehensive analysis of the effect that different choices of material for the surface have on the energy shift.
 
\subsection{Intermediate trap frequency: $\frac{|e| B_0}{m} \ll \omegatrap  \ll \frac{c}{|z|}$}

For $\omegatrap \gg \frac{|e| B_0}{m}$ an electron bound to an atom is the most relevant physical system. But the additional condition $\omegatrap  \ll \frac{c}{|z|}$ would for the atomic `trap' frequency of $10^{15}$Hz constrain the distance to $|z| \ll 100 \text{nm}$, which is quite unrealistic in practice. Nevertheless, to provide an estimate we note that  $\frac{|e| B_0}{m} \ll \omegatrap$ leads to $\Omega$, $\Delta_R$, and $\Delta_L$ all being roughly equal to $\omegatrap$, so that $\omegatrap  \ll \frac{c}{|z|}$ implies $|z|\Delta_i\ll 1$. Therefore, the small-argument asymptotics in Eq.~\eqref{smalltheta} applies and Eq.~\eqref{perfectShift} together with $\Delta_i/\Omega \approx 1-h_i eB_0/(2m\omega_H)$ gives the leading term of the Zeeman energy shift of an electron near a perfect reflector as
\begin{equation}
\Delta \mathcal{E} \approx -\frac{3 e^3\sigma_z B_0}{256\pi m^3 \omegatrap z^3} \left(\nu_R+\nu_L+1 \right)\;.
\end{equation}
For an arbitrary dielectric the energy shift in this regime is awkward to analyse because small $|z|$ means poor convergence in the $k_z$ integrals in Eqs.~\eqref{DeltaEDipRenormExplicitDeformed}--\eqref{DeltaESpinFinal}. We skip a more detailed discussion because of the lack of realistic applicability of this regime, as noted above.

\subsection{Large trap frequency: $\frac{|e| B_0}{m} \ll \frac{c}{|z|} \ll \omegatrap $}

In this case the most relevant physical system is again an atomic electron. 
The constraint $\omegatrap  \gg \frac{c}{|z|}$ now corresponds to large distances $|z| \gg 100 \text{nm}$. On the other hand, an upper limit on the distance is imposed by the weak-field constraint $\frac{|e| B_0}{m} \ll \frac{c}{|z|} $, which corresponds to $|z|\ll 3$mm, as discussed earlier. But the range $100\text{nm} \ll |z| \ll 3\text{mm}$ is realistically accessible by experiments.
For this reason we shall from now on focus in this third asymptotic regime, $\frac{|e| B_0}{m} \ll \frac{c}{|z|} \ll \omegatrap$, and analyse the expressions for the energy shift in Eqs.~\eqref{DeltaEDipRenormExplicitDeformed}--\eqref{DeltaESpinFinal} in more detail for this case. We note that $\Omega$, $\Delta_R$, and $\Delta_L$ are again all roughly equal to $\omegatrap$, so that $\frac{c}{|z|} \ll \omega_H$ implies $\Delta_i  |z|\gg 1$ in natural units.

\section{Asymptotic shift for large distances and a weak field}
\subsection{Transitions between Landau levels}
To evaluate the energy shifts due to virtual transitions between Landau levels, given by Eqs.~\eqref{DeltaEDipRenormExplicitDeformed} and \eqref{DeltaEQuadRenormExplicitDeformed}, we again split the integrals into contributions from the path $C'$, and those from $C_\Delta$ as shown in Fig.~\ref{ContourDielectricHBE},
\begin{align*}
\Delta  \mathcal{E}_2^E &=\Delta  \mathcal{E}_2^E(C_\Delta )+\Delta  \mathcal{E}_2^E(C')\\
\Delta \mathcal{E}_2^{Q,1}&= \Delta \mathcal{E}_2^{Q,1}(C_\Delta)+\Delta \mathcal{E}_2^{Q,1}(C')\;.
\end{align*}
As in Sec.~\ref{sec:perfectrefl}, we evaluate the contributions from the path $C_\Delta$ by first calculating the residue and then changing variables in the remaining $k_\parallel$ integrals according to Eq.~\eqref{substitutions}. Defining the abbreviation
\begin{align}
\zeta_i \equiv -\Delta_i z  \; ,
\end{align}
we find 
\begin{align}
\Delta \mathcal{E}_2^E =&  \frac{e^2\sigma_z}{32\pi m^2\Omega} \sum_i h_i \Delta_i^4 \nu_i  \notag\\
&\times \Bigg\{\int_0^1 dx \left[ R_{\text{TE}}^+(x) -x^2 R_{\text{TM}}^+(x) \right]\sin({2\zeta_i x}) \notag\\
&- \re \int_0^\infty dy \left[ R_{\text{TE}}^-(y)+y^2R_{\text{TM}}^-(y) \right] e^{-2\zeta_i y} \Bigg\}\label{DeltaEPoleContributions}
\end{align}
and
\begin{align}
\Delta \mathcal{E}_2^{Q,1} =& -\frac{e^2\sigma_z}{16\pi m^2\Omega} \sum_i h_i \Delta_i^4 \nu_i \notag\\
&\times \Bigg\{ \int_0^1 dx\;(x^2-1) R_{\text{TE}}^+(x) \sin({2\zeta_i x})\notag\\
&+ \re \int_0^\infty dy\  (y^2+1) R_{\text{TE}}^-(y) e^{-2\zeta_i y}\Bigg\}\label{DeltaEQuadContributions}
\end{align}
where
\begin{align*}
R^\pm_{\text{TE}}(\alpha) &= \frac{\alpha - \sqrt{ \alpha^2 \pm  (n^2-1)}}{\alpha + \sqrt{ \alpha^2 \pm  (n^2-1)}} \\
R^\pm_{\text{TM}}(\alpha) &= \frac{n^2\alpha - \sqrt{ \alpha^2 \pm  (n^2-1)}}{n^2\alpha + \sqrt{ \alpha^2 \pm  (n^2-1)}} \;.
\end{align*}
We now have explicit expressions for the entire contribution from the path $C_\Delta$ to terms that originate from transitions between Landau levels,
\begin{equation}
\Delta  \mathcal{E}_D(C_\Delta )=\Delta  \mathcal{E}_2^E(C_\Delta )+\Delta  \mathcal{E}_2^{Q,1}(C_\Delta )\end{equation}
The integrals in Eqs.~\eqref{DeltaEPoleContributions} and \eqref{DeltaEQuadContributions} converge rapidly for large $\zeta_i$, and their asymptotic analysis is straightforward to derive. For the $x$ integrals, which came from $k_\parallel < \Delta_i$, this is achieved via repeated integration by parts, and for the $y$ integrals, which came from $k_\parallel > \Delta_i$, one applies Watson's lemma and Taylor expands the integrand for small $y$. In total the sum of the large $\zeta_i$ asymptotics of Eqs.~\eqref{DeltaEPoleContributions} and \eqref{DeltaEQuadContributions} is%
\begin{align}
\Delta  \mathcal{E}_D(C_\Delta )(\zeta_i \gg 1)=& \frac{e^2\sigma_z}{32\pi m^2\Omega z^2}\frac{n-1}{n+1}\sum_i h_i \Delta_i^2 \nu_i \notag \\
& \times \bigg[  \zeta_i\cos(2\zeta_i)+\mathcal{O}(\zeta_i^0)\bigg] \label{DeltaEPolesFinal}
\end{align}
The contribution along $C'$,
\begin{align}
\Delta  \mathcal{E}_D(C' )(\zeta_i \gg 1)=\Delta  \mathcal{E}_2^E(C' )+\Delta  \mathcal{E}_2^{Q,1}(C' )\;,
\end{align}
can also be calculated but gives lengthy expressions. We do not quote those here because their order of magnitude is quite easily seen from the large $\zeta_i\equiv -z\Delta_i$ expansion of Eq.~\eqref{perfectCprime}, which can be read off from Eq.~\eqref{largetheta} with $\nu_i=0$,
\begin{align}
\Delta \mathcal{E}_D(C')(\zeta_i \gg 1)\sim & -\frac{e^2\sigma_z}{128\pi^2 m^2\Omega z^3} \sum_i h_i \Delta_i \notag\\
&\times \left[ -\frac{4}{\zeta_i} + O\left(\zeta_i^{-3}) \right) \right]
\end{align}
This is negligible compared to Eq.~\eqref{DeltaEPolesFinal}.
Thus the large $\zeta_i$ asymptotics of the energy shift $\Delta  \mathcal{E}_D$ due to transitions between Landau levels is dominated by the contributions from $C_\Delta$
and given by Eq.~\eqref{DeltaEPolesFinal}. 

\subsection{Transitions within the same Landau level}

Turning our attention to the energy shifts due to transitions within the same Landau level \eqref{DeltaEFirstOrderFinal} and \eqref{DeltaESpinFinal}, we note these are independent of $\Delta_i$,
and entirely the same as for an electron not bound in a trap, which is the system that has been investigated in detail in Ref.~\cite{MagMomPRA}. The shift can be calculated for dielectric as well as conducting surfaces. Here we only note that in all cases it is of the same order of magnitude as that for the perfect reflector, given in Eq.~\eqref{perfectSpinflip} and thus also negligible compared to Eq.~\eqref{DeltaEPolesFinal}.

\section{Discussion}
We have seen that for $\frac{|e| B_0}{m} \ll \frac{c}{|z|} \ll \omegatrap $ the energy shift is dominated by the large $\zeta_i$ asymptotics 
of contributions due to transitions between Landau levels and just those from the path $C_\Delta$. Substituting back the definition of $\zeta_i$ and writing out the sum over $i$, the leading term of Eq.~\eqref{DeltaEPolesFinal} is 
\begin{align}
\Delta \mathcal{E}(\zeta_i \gg 1)=\frac{e^2\sigma_z}{32\pi m^2 z\Omega}\; \frac{n-1}{n+1}&\bigg[  \nu_R  {\Delta_R^3} \cos(2\Delta_R z)\notag \\
&\!\!\!\!-\nu_L  {\Delta_L^3}{}   \cos(2\Delta_L z)\bigg] \label{HBEFinalResultExpanded}\end{align}
In analogy to the case of an electron without a trap \cite{MagMomPRA}, one might want to extract a magnetic moment from this shift by isolating the coefficient of $B_0$ as $B_0\to 0$ in accordance with to $\Delta\mathcal{E} = -\Delta\mu\  \sigma_z B_0$. Expanding Eq.~\eqref{HBEFinalResultExpanded} for small $B_0$ and extracting the coefficient of $\sigma_z B_0$, we find for the magnetic moment shift the rather surprising result
 \begin{align}
\Delta \mu (|\omegatrap z| \gg 1) =&-\frac{e^3 (\nu_L+\nu_R)  }{64\pi m^3   } \; \frac{n-1}{n+1}\; \omegatrap\notag \\
&\times \left[2  \omegatrap\sin(2 \omegatrap z )-\frac{3\cos(2  \omegatrap z )}{z}\right]\label{HBEMagMomStrangeResult}
\end{align}
This is of course unphysical in that it oscillates with undiminished amplitude as $|z|\to\infty$. 
To track down the source of this perplexity, we define the abbreviation $\Lambda \equiv -e B_0/2m>0$, turning Eqs.~\eqref{OmegaDef} and \eqref{DeltaDef} into
\begin{align}
\Omega &= \sqrt{\omegatrap^2+\Lambda^2} & \Delta_i = \sqrt{\omegatrap^2 + \Lambda^2}+h_i \Lambda	 \; , \label{LambdaDefs}
\end{align}
and look at the behaviour of the relevant part of Eq.~\eqref{HBEFinalResultExpanded}, namely
\begin{align}
\frac{\Delta_i^3}{z\Omega}\cos(2\Delta_i z) =&\frac{ \left(\sqrt{\omegatrap^2 + \Lambda^2}+h_i \Lambda\right)^3 }{z \sqrt{\omegatrap^2+\Lambda^2} }\notag \\
&\times\cos\left[2\left(\sqrt{\omegatrap^2 + \Lambda^2}+h_i \Lambda \right) z\right] \label{CosTermTermsofLambda} 
\end{align}
As $\Lambda\to 0$ the amplitude behaves as expected and vanishes with $z^{-1}$ for large $|z|$.
However, $\Lambda\to 0$ also causes the wavelength of the cosine to change; Taylor expansion in this limit gives
\begin{align*}
\cos\bigg[2 & \left(\sqrt{\omegatrap^2 + \Lambda^2} +h_i \Lambda \right) z\bigg]\\
&\stackrel{\Lambda\rightarrow 0}{\longrightarrow} \  \cos 2\omega_H z -2 h_i z \Lambda \sin 2\omega_H z +O(\Lambda^2)\;,
\end{align*}
so that one picks up a factor of $z$ when selecting the term linear in $\Lambda$ in this expansion. Combined with the amplitude this then leads to the $z^0$ dependence of the magnetic moment in Eq.~\eqref{HBEMagMomStrangeResult}. 

A similar dependence on the magnetic field $B_0$ was found in Ref.~\cite{BartonFawcett} for the energy shift in the `retarded' regime $\frac{|e| B_0}{m} \gg \frac{c}{|z|} $ of a free electron near a perfect reflector. We are in the weak-field or `non-retarded' regime $\frac{|e| B_0}{m} \ll \frac{c}{|z|}$, but simultaneously in a retarded regime with respect to the trap frequency $\omegatrap$ and the distance $z$, since we have $\omegatrap \gg \frac{c}{|z|}$. In other words, retardation matters here because, during the time it takes for a photon to make a round trip from the electron to the interface and back, the electron's state in the trap has evolved significantly. The phase of that state is important, which is why we get oscillatory terms. Magnetic moments are strictly defined only in the non-retarded regime, so what we are seeing in Eq.~\eqref{HBEMagMomStrangeResult} is just an indication that it is not sensible to consider the magnetic moment for a trapped electron with $\omegatrap \gg \frac{c}{|z|}$. Therefore the Zeeman energy shift \eqref{HBEFinalResultExpanded} is our final result.

If we track down from which parts of the various integrals the terms in Eq.~\eqref{HBEFinalResultExpanded} have arisen, we see that they have come from photons with frequency $\omega=\Delta_i$, i.e.~resonances with transitions between either left- or right-circular Landau states, and among those from photons with $k_\|=0$, i.e.~photons that are incident and reflected normal to the surface. Mathematically they came from a residue around $\omega=\Delta_i\approx\omega_H$, which indicates that for a general dispersive dielectric or conductor one gets the same expression as in Eq.~\eqref{HBEFinalResultExpanded} but with the refractive index $n$ at the frequency of this resonance, 
$$
n \to \sqrt{\epsilon(\omega\approx\omega_H)} \;.
$$
With that replacement the leading term of the Zeeman energy shift for a dispersive medium is also given by Eq.~\eqref{HBEFinalResultExpanded}. A potentially interesting variation would arise if the refractive index of surface has an absorption resonance and an area of anomalous dispersion lying between $\Delta_R$ and $\Delta_L$, when the two cosine terms in Eq.~\eqref{HBEFinalResultExpanded} could each have very different pre-factors $\frac{n-1}{n+1}$.

Another interesting observation concerns the comparison of the general asymptotic result in Eq.~\eqref{HBEFinalResultExpanded} with the result in Eq.~\eqref{perfectShift} for a perfectly reflecting plane, for which all integrals could be calculated exactly for any distance: as shown by Eq.~\eqref{largetheta}, the leading term in the large-distance limit of Eq.~\eqref{perfectShift} does agree with the perfect-reflector limit $n\to\infty$ of Eq.~\eqref{HBEFinalResultExpanded}. This is in sharp contrast to the Zeeman shift and magnetic moment of an electron without any trap, as discussed in detail in Ref.~\cite{MagMomPRA}. This is because extreme long-wavelengths excitations play a crucial role in the weak-field Zeeman shift. For a free electron these such excitations reach right down to zero frequency, where conductors and insulators behave very differently --- and hence the Zeeman shift comes out very different, but for a trapped electron with $\frac{|e| B_0}{m} \ll  \omegatrap $ the excitation spectrum has a lower cut-off due to left- and right-circular Landau transitions both requiring an energy of about $\omega_H$, and hence what matters is the refractive index of the material at that frequency but not whether the material is a conductor or an insulator, i.e.~whether the polarizability of the material diverges in the static limit or not.

For experiments one is interested not so much in the Zeeman energy level but in the splitting between spin-up and spin-down states, which can be probed by looking for spin-flip resonances. 
We can extract from Eq.~\eqref{HBEFinalResultExpanded} an expression for the shift $\delta$ in the spin energy splitting and express this in units of the unperturbed Zeeman spin energy level splitting $\delta_0=|e| B_0/m$
\begin{align}
\frac{\delta (\zeta_i \gg 1)}{\delta_0}=&\frac{|e|}{16\pi m zB_0 \Omega}\frac{n-1}{n+1}\bigg[  \nu_R  {\Delta_R^3} \cos(2\Delta_R z)\notag \\
&\qquad\quad -\nu_L  {\Delta_L^3}{}   \cos(2\Delta_L z)\bigg] \label{HBEFinalResultExpandedRatio} \end{align}
Since the quantities $\Delta_i$ and $\Omega$ are frequencies, Eq.~\eqref{HBEFinalResultExpandedRatio} reads in S.I. units
\begin{align}
&\frac{\delta (\zeta_i \gg 1)}{\delta_0}=\frac{\hbar}{4\pi\epsilon_0 c^4} \frac{|e|}{4m zB_0 \Omega}\frac{n-1}{n+1}\notag \\
& \qquad \times \bigg[  \nu_R  {\Delta_R^3} \cos(2\Delta_R z/c)-\nu_L  {\Delta_L^3}{}   \cos(2\Delta_L z/c)\bigg] \label{HBERatioRealUnits}\end{align}
As discussed in section \ref{AsymptoticRegimesHBE}, parameters which are consistent with our choice of asymptotic regime are
\begin{align}
B_0 &\sim \text{T}  & |z| &\sim 10 \mu \text{m} & \omega_H &\sim  10^{15} \text{Hz}
\end{align}
for which  $\Delta_L |z|/c\approx \Delta_R |z|/c\approx 30$, meaning that we are at the low end of the region $\Delta_i |z|/c \gg 1$. Nevertheless, we substitute the values for $B_0$ and $\omegatrap$ into Eq.~\eqref{HBERatioRealUnits} to find the $z$ dependence of the size of the shift for distances satisfying $ 0.1 \mu \text{m} \ll |z| \ll 3 \text{mm} $

\begin{equation*}
\left| \frac{\delta(\zeta_i \gg 1)}{\delta_0} \right| \approx  (\nu_R-\nu_L)\frac{n-1}{n+1 }\cdot \frac{10^{-11}\mu \text{m}}{|z|} \cos(6 \mu \text{m}^{-1} z) 
\end{equation*}
where $z$ is measured in $\mu$m. Taking somewhat optimistically $(\nu_R - \nu_L)\frac{n-1}{n+1 }\approx 10$ \cite{fn3}
and a distance $|z|$ of $10\mu$m gives for the amplitude of the shift
\begin{align}
\left|\frac{\delta(\zeta_i \gg 1)}{\delta_0}\right| \approx 10^{-11} \; , \end{align}
which is very small. Currently the best bound-state magnetic moment measurements reach down to an accuracy of about $10^{-11}$ \cite{Werth2013}, so measurement of the shift in Eq.~\eqref{HBERatioRealUnits} is right on the edge of experimental viability.

\section{Summary and conclusions}

We have derived an integral formula, Eq.~\eqref{EnergyShiftSum}, that gives the shift in the difference between the spin energy levels of an electron trapped near a surface. We have evaluated our formula for the most relevant orders of magnitude of the physical parameters of the system. The shift is either essentially the same as for an untrapped electron investigated in Ref.~\cite{MagMomPRA}, or its leading behaviour is oscillatory for excited states and given by Eq.~\eqref{HBEFinalResultExpanded} with the refractive index $n$ at the trap frequency.
We have shown that this oscillatory energy shift is small, but possibly not so far beyond the reach of current experiments that this effect could not come within the reach of Zeeman shift measurements in the near future. 

\section*{Acknowledgment}
It is a pleasure to thank Jos\'e Verd\'u for advice. Financial support from the UK Engineering and Physical Sciences Research Council (EPSRC) is gratefully acknowledged.

\appendix
\section{Schr\"{o}dinger eigenstates of an electron subject to confinement by a constant magnetic field and a harmonic potential}\label{HOscApp}
We require the Schr\"{o}dinger eigenstates for the electronic part of the Hamiltonian \eqref{H0}, which is
\begin{align*}
H^H_e &= \frac{\left(\hat{p}_x+\frac{eB_0}{2}\hat{y}\right)^2}{2m}+\frac{\left(\hat{p}_y-\frac{eB_0}{2}\hat{x}\right)^2}{2m}\\
&\quad +\frac{\hat{p}_z^2}{2m}+\frac{m \omegatrap^2}{2}(\hat{x}^2+\hat{y}^2)\\
&= \frac{\hat{p}_x^2+\hat{p}_y^2+\hat{p}_z^2}{2m}+\frac{m\Omega^2}{2}(\hat{x}^2+\hat{y}^2) - \frac{eB_0}{2m}\hat{L}_z
\end{align*}
with the definitions
\begin{align*}
\Omega^2 &=\omegatrap^2+\left(\frac{eB_0}{2m}\right)^2\\
\hat{L}_z  &=\hat{x}\hat{p}_y-\hat{p}_x\hat{y}
\end{align*}
Introducing the operators
\begin{subequations}
\label{PositionAndMomentumBound}
\begin{align}
\hat{x}&=\frac{1}{\sqrt{2 m \Omega}}(\ax+\axd)\label{xypxpyDefs1}\\
\hat{y}&=\frac{1}{\sqrt{2 m \Omega}}(\ay+\ayd)\\
\hat{p}_x&=i \sqrt{\frac{m\Omega}{2}} (\axd-\ax)\\
\hat{p}_y&=i \sqrt{\frac{m\Omega}{2}} (\ayd-\ay)\label{xypxpyDefs4}
\end{align}
\end{subequations}
the Hamiltonian may be written
\begin{align}
H^H_e = \frac{\Omega}{2}(\axd\ax+\ax\axd &+\ayd\ay+\ay\ayd)\notag \\
&-\frac{i e B_0}{2m}(\ax\ayd- \ay\axd)
\end{align}
Further defining the operators for right and left-circular quanta
\begin{align}
\aR &=\frac{1}{\sqrt{2}}\left(\ax-i \ay \right) \label{bR}\\
\aL &=\frac{1}{\sqrt{2}}\left(\ax+i \ay\right)  \label{bL}
\end{align}
one finds:
\begin{align}
H^H_e= \frac{\Omega}{2}(\aRd \aR+\aR \aRd &+\aLd \aL+\aL \aLd )\notag \\
&+\frac{ e B_0}{2m}(\aL \aLd - \aR \aRd)
\end{align}
Taking advantage of the commutation relation
\begin{equation}
[\aR,\aRd] = 1 =[\aL,\aLd] ,
\end{equation}
this can be written as
\begin{equation}
H^H_e=\left(\Omega-\frac{eB_0}{2m}\right)\aRd \aR+\left(\Omega+\frac{eB_0}{2m}\right)\aLd \aL + \Omega
\end{equation}
Since our $e<0$, the limit $\omegatrap \to 0$ is equivalent to the limit $\Omega \to -\frac{eB_0}{2m}$. In this limit, the above Hamiltonian becomes
\begin{equation}
H^H_e(\omegatrap \to 0)=-\frac{eB_0}{m}\left(\aRd \aR+\frac{1}{2}\right) \label{HHnotrap}
\end{equation}
which is the usual statement of the Landau-quantized Hamiltonian, and shows infinite degeneracy in the left-circular quanta. By contrast, the energy eigenvalues of $H_e^H$ for a state $\ket{\nu_L,\nu_R}$ are
\begin{equation}
E_{\nu_R,\nu_L}=\left(\Omega-\frac{e B_0}{2m}\right)\nu_R+\left(\Omega+\frac{e B_0}{2m}\right)\nu_L+\Omega \label{EnLnR}
\end{equation}
where $\nu_L$ and $\nu_R$ are eigenvalues of the number operators for left- and right-circular quanta, $\nu_i \ket{i} = \aid \ai \ket{i}$. Using the definition \eqref{DeltaDef} the Hamiltonian may be written as:
\begin{equation}
H^H_e = \Delta_L b_L^\dagger b_L +\Delta_R b_R^\dagger b_R + \Omega \label{HamiltonianDelta}
\end{equation}
The canonical momenta can be written in terms of $\aR$ and $\aL$ via eqs.~(\ref{xypxpyDefs1}-\ref{xypxpyDefs4})

\begin{subequations}
\begin{align}
\hat{\pi}_x &= \hat{p}_x+\frac{e B_0}{2}\hat{y}\notag \\
&= \frac{i}{2}\sqrt{\frac{m}{\Omega}}\left[\Delta_R (\aRd-\aR)+\Delta_L (\aLd-\aL)\right]\\
\hat{\pi}_y &= \hat{p}_y-\frac{e B_0}{2}\hat{x}\notag \\
&=  \frac{1}{2}\sqrt{\frac{m}{\Omega}}\left[\Delta_R(\aRd+\aR)-\Delta_L(\aLd+\aL)\right]
\end{align}
\end{subequations}
and of course $\hat{\pi}_z = \hat{p}_z$.  These equations for the canonical momenta show that their action on a state of definite $\nu_L$ and $\nu_R$ can change either $\nu_L$ or $\nu_R$ but not both; their non-zero matrix elements are
\begin{align}
\bra{\nu_i+1, \nu_j} \hat{\pi}_x \ket{\nu_i, \nu_j} &=   \frac{i}{2} \sqrt{\frac{m}{\Omega}} \Delta_i \sqrt{\nu_i+1}\notag\\
\bra{\nu_i-1, \nu_j} \hat{\pi}_x \ket{\nu_i, \nu_j} &= - \frac{i}{2} \sqrt{\frac{m}{\Omega}} \Delta_i \sqrt{\nu_i}\notag\\
\bra{\nu_i+1, \nu_j} \hat{\pi}_y \ket{\nu_i, \nu_j} &= h_i \sqrt{\frac{m}{\Omega}} \Delta_i \sqrt{\nu_i+1}\notag\\
\bra{\nu_i-1, \nu_j} \hat{\pi}_y \ket{\nu_i, \nu_j} &= h_i \sqrt{\frac{m}{\Omega}}  \Delta_i \sqrt{\nu_i}
\end{align}
where definition \eqref{hDef} has been used. It is also useful to have the matrix element of the displacement operator in the directions parallel to the surface:
\begin{align}
\bra{\nu_i+1, \nu_j} (x-x_0) \ket{\nu_i, \nu_j} &= \frac{1}{2\sqrt{m\Omega}} \sqrt{\nu_i+1}\notag\\
\bra{\nu_i-1, \nu_j} (x-x_0)\ket{\nu_i, \nu_j} &= \frac{1}{2\sqrt{m\Omega}}  \sqrt{\nu_i}\notag\\
\bra{\nu_i+1, \nu_j}( y-y_0) \ket{\nu_i, \nu_j} &= -h_i\frac{i}{2\sqrt{m\Omega}} \sqrt{\nu_i+1}\notag\\
\bra{\nu_i-1, \nu_j} (y-y_0) \ket{\nu_i, \nu_j} &= h_i\frac{i}{2\sqrt{m\Omega}} \sqrt{\nu_i} \label{PositionMatrixElements}
\end{align}

\section{Normal modes for the non-dispersive dielectric}\label{ModesAppendix}
We consider a semi-infinite slab of non-magnetic, non-dispersive material that fills the half-space $z>0$ as shown in Fig.~(\ref{HalfSpace}). The dielectric function is:
\begin{equation}
\epsilon({\bf r})=n^2({\bf r}) = 1+\Theta(z)(n^2-1)
\end{equation}
where $n^2\geq 1$ is the index of refraction and independent of frequency. Following \cite{CarnigliaMandel, PRDEberleinRobaschik,MagMomPRA} we use the electromagnetic boundary conditions at the interface of two non-magnetic materials  to derive a mode expansion for the electromagnetic field in this geometry. We denote wave vectors that exist on the vacuum side as $\mathbf{k}$, and those on the medium side as $\mathbf{k}^d$. We further decompose these into components parallel to the surface ($\mathbf{k_\parallel}$) and perpendicular to it ($k_z$). The modes are labelled by the region they are incident from (vacuum or medium), momentum $\mathbf{k}$ and polarization $\lambda$, and are separated into incident, reflected and transmitted parts. A superscript $R$ denotes a reflected $\bf{k}$ vector, with the same $\mathbf{k_\parallel}$ but opposite $k_z$ to the incident wave. The modes are subject to the constraint that sgn$(k_z) =$ sgn$(k_z^d)$ which ensures that the transmitted parts of a mode move in the same direction as their incident part. Modes that are incident from the dielectric may suffer total internal reflection, and thus be evanescent on the vacuum side. This corresponds to a certain range of values for $k_z^d$ resulting in imaginary $k_z$. The modes are: 
\begin{widetext}
\begin{align}
\B{f}_{\B{k} \lambda}^{\text{vac}} &= \frac{1}{(2\pi)^{3/2}} \left\{\Theta(-z)[e^{i\mathbf{k}\cdot \mathbf{r}} \hat{\B{e}}_\lambda(\mathbf{k})+R^\text{vac}_\lambda e^{i\mathbf{k}^R\cdot \mathbf{r}} \hat{\B{e}}_\lambda(\mathbf{k}^R)] + \Theta(z)T^\text{vac}_\lambda e^{i\mathbf{k}^d\cdot \mathbf{r}} \hat{\B{e}}_\lambda(\mathbf{k}^d) \right\} \notag \\
\B{f}_{\B{k}\lambda}^{\text{med}} &=\frac{1}{(2\pi)^{3/2}} \frac{1}{n}\left\{\Theta(z)[e^{i\mathbf{k}^d\cdot \mathbf{r}} \hat{\B{e}}_\lambda(\mathbf{k}^d)+R^\text{med}_\lambda e^{i\mathbf{k}^{dR}\cdot \mathbf{r}} \hat{\B{e}}_\lambda(\mathbf{k}^{dR})] + \Theta(-z)T^\text{med}_\lambda e^{i\mathbf{k}\cdot \mathbf{r}} \hat{\B{e}}_\lambda(\mathbf{k}) \right\}   \label{NonDispModes}
\end{align}
\end{widetext}
where the $\hat{\B{e}}_\lambda (\mathbf{k})$ are unit polarization vectors obeying $\mathbf{k} \cdot \hat{\B{e}}_\lambda(\mathbf{k})=0$. A convenient choice is:
\begin{subequations}
\label{PolVecs}
\begin{align}
\hat{\mathbf{e}}_\text{TE}(\mathbf{k}) &= \frac{1}{k_\parallel}\left( k_y,-k_x,0\right)\\
\hat{\mathbf{e}}_\text{TM}(\mathbf{k})  &= \frac{1}{k k_\parallel}\left( k_xk_z,k_yk_z,-k_\parallel^2 \right)
\end{align}
\end{subequations}
The reflection and transmission coefficients are given by the standard Fresnel expressions
\begin{align}
R^\text{vac}_{TE} &= \frac{k_z - k_z^d}{k_z + k_z^d}  &T^\text{vac}_{TE} &= \frac{2k_z}{k_z + k_z^d} \notag  \\
R^\text{vac}_{TM} &= \frac{n^2 k_z - k_z^d}{n^2 k_z + k_z^d}  &T^\text{vac}_{TM} &= \frac{2n k_z}{n^2 k_z + k_z^d} \notag \\
R^\text{med}_\lambda &= - R^v_\lambda  &T^\text{med}_\lambda &= \frac{k_z^d}{k_z}T_\lambda^v 
\end{align}
with 
\begin{equation}
k_z^d = \sqrt{n^2(k_z^2+k_\parallel^2)-k_\parallel^2}\, .
\end{equation}
These modes are the same as those used in \cite{EberleinRobaschik} (where the authors use slightly different conventions as to where to put factors of $n$). They are normalized according to
\begin{equation}
\int d^3 \B{r} \; n^2(\B{r}) \B{f}_{\B{k}\lambda}(\B{r}) \cdot \B{f}_{\B{k}'\lambda'} (\B{r})= \delta_{\lambda \lambda'} \delta^{(3)}(\B{k}-\B{k}')
\label{normf}
\end{equation}
for both $\B{f}_{\B{k} \lambda}^{\text{vac}}$ and $\B{f}_{\B{k}\lambda}^{\text{med}} $, which ensures that the radiation Hamiltonian appears in the canonical form (\ref{Hrad}).


\begin{thebibliography}{25}
\bibitem [{fn1()}]{fn1}%
  \BibitemOpen
  \href@noop {} {}\bibinfo {note} {Note that the mode functions
  $\mathbf{f}_{\mathbf{k}\lambda}$ differ by a factor $\sqrt{2\omega}$ from
  those used in \cite{MagMomPRA,BennettEberleinMagMomNJP}.}\BibitemShut {Stop}%
\bibitem [{fn3()}]{fn3}%
  \BibitemOpen
  \href@noop {} {}\bibinfo {note} {The $z$ component of the orbital angular
  momentum, $\hat{L}_z = \hat{x}\hat{p}_y - \hat{p}_x \hat{y}$, expressed in
  terms of annihilation and creation operators of right- and left-circular
  Landau quanta is $ \hat{L}_z =b_R^\dagger b_R-b_L^\dagger b_L$, as easily
  shown from Eqs.~(\ref{PositionAndMomentumBound}), (\ref{bR}), and (\ref{bL}).
  Thus $|\nu_R,\nu_L\rangle$ are eigenstates of $\hat{L}_z$ with eigenvalue
  $\nu_R-\nu_L$.}\BibitemShut {Stop}%
\bibitem [{fn2()}]{fn2}%
  \BibitemOpen
  \href@noop {} {}\bibinfo {note} {We shall see in Section \ref{Sec:Qzero} that
  $\Delta \mathcal{E}^{\text{q},2}_2$ is in fact zero regardless of the value
  of $B_0$.}\BibitemShut {Stop}%
\end{thebibliography}
%

\end{document}